%% file: template.tex
\documentclass{article}
\usepackage{authblk}

\usepackage{arxiv}
\usepackage{tabularx}
\usepackage[utf8]{inputenc} 
\usepackage[T1]{fontenc}    
\usepackage{hyperref}       
\usepackage{url}            
\usepackage{booktabs}       
\usepackage{amsfonts}       
\usepackage{nicefrac}       
\usepackage{microtype}      
\usepackage{lipsum}
\usepackage{graphicx}
\usepackage{subcaption}
\usepackage[T1]{fontenc}
\usepackage[utf8]{inputenc}
\graphicspath{ {./images/} }
\usepackage[most]{tcolorbox}
\DeclareUnicodeCharacter{00ED}{\'{\i}} 
\DeclareUnicodeCharacter{0301}{'}      

\newcommand{\etal}{et al.\ }
\newtcolorbox{SummaryBox}[1]{enhanced,arc=1mm,outer arc=1mm,
  boxrule=0mm,toprule=0mm,bottomrule=0mm,left=1mm,right=1mm,leftrule=2pt,
  titlerule=0mm,toptitle=0mm,bottomtitle=0mm,top=0mm,
  colframe=blue!50!black,colback=blue!5!white,coltitle=blue!50!black,
  colbacktitle=yellow!50!white,colback=green!5!white,
  title=#1,
  fonttitle=\bfseries\sffamily\normalsize,fontupper=\normalsize\itshape,
}

\title{What's Privacy Good for? \\Measuring Privacy as a Shield from Harms due to Personal Data Use}

\author{
 Sri Harsha Gajavalli \\
  School of Coumputing and Augmented Intelligence\\
  Arizona State University\\
  Tempe, AZ 85281 \\
  \texttt{gsh@asu.edu} \\
   \and
 Junichi Koizumi \\
  School of Coumputing and Augmented Intelligence\\
  Arizona State University\\
  Tempe, AZ 85281 \\
  \texttt{jkoizumi1@asu.edu} \\
  \and
 Rakibul Hasan \\
  School of Coumputing and Augmented Intelligence\\
  Arizona State University\\
  Tempe, AZ 85281 \\
  \texttt{rakibul.hasan@asu.edu} \\

}

\begin{document}
\maketitle
\include{sections/0_abstract}

\input{sections/1_introduction}
\input{sections/2_literature}
\input{sections/3_methodology}
\input{sections/4_results}
\input{sections/5_discussions}
\input{sections/appendix}

\bibliographystyle{unsrt}  
\bibliography{references, references-rakib}

\end{document}

%% file: sections/0_abstract.tex
\begin{abstract}

We propose a harm-centric conceptualization of privacy that asks: What harms from personal data use can privacy prevent? The motivation behind this research is limitations in existing privacy frameworks (e.g., Contextual Integrity) to capture or categorize many of the harms that arise from modern technology's use of personal data. We operationalize this conceptualization in an online study with 400 college and university students. Study participants indicated their perceptions of different harms (e.g., manipulation, discrimination, and harassment) that may arise when artificial intelligence-based algorithms infer personal data (e.g., demographics, personality traits, and cognitive disability) and use it to identify students who are likely to drop out of a course or the best job candidate. The study includes 14 harms and six types of personal data selected based on an extensive literature review.

Comprehensive statistical analyses of the study data show that the 14 harms are internally consistent and collectively represent a general notion of privacy harms. The study data also surfaces nuanced perceptions of harms, both across the contexts and participants' demographic factors. Based on these results, we discuss how privacy can be improved equitably. Thus, this research not only contributes to enhancing the understanding of privacy as a concept but also provides practical guidance to improve privacy in the context of education and employment.



\end{abstract}

%% file: sections/1_introduction.tex
\section{Introduction}
A crucial goal of usable privacy research has been to define and measure privacy. Earlier efforts centered around the concept of disclosure (or concealment) of private data, where privacy is violated when such data is leaked~\cite{xuExaminingFormationIndividuals-2008}. However, categorizing data as private (versus public) is arbitrary and malleable~\cite{soloveDataWhatDataDoes2024}, their disclosure is often a necessity, and they can be \textit{inferred} using non-private data. Research on defining and measuring related (or proxy) concepts, such as privacy concerns~\cite{dinevExtendedPrivacyCalculusECommerce2006}, primarily to understand and predict data-disclosing behaviors, remained unsatisfactory, with the coining of (now-debunked~\cite{soloveMythofPrivacyParadox2021}) terms like privacy-paradox to explain concern-behavior mismatches. Likewise, framing privacy as \textit{control over personal data} led to measures like notice and consent, which ultimately degraded privacy~\cite{solove_murky_2023}. 


A different perspective, privacy as contextual integrity (CI)~\cite{nissenbaum2004CI}, sidesteps defining data as inherently private or public; instead, it declares a data flow as a privacy violation if it is misaligned with established norms or expectations in a given context. CI overcomes many of the limitations mentioned above, better explains people's privacy-related attitudes, and has gained wide popularity in the last two decades within the privacy research community~\cite{kumarRoadmapApplyingCI2024}. CI, however, only concerns semantically meaningful data (e.g., age); the exclusion of information that can be inferred from other data renders it incapable of determining privacy violations for a large number of digital services based on AI (artificial intelligence) that constantly perform such inferences. Moreover, CI's reliance on norms prevents it from recognizing novel privacy violations from established data-disclosing practices.



In this paper, we conceptualize privacy as a \textit{shield against harms caused by the use of personal data in a certain context}. The fact that privacy is much more than data protection~\cite{angel2024distinguishingPrivacy}, and its role in achieving other human values, including freedom, protection from manipulation, and identity building, has long been recognized~\cite {richardsWhyPrivacyMatters-2021}. Lack of privacy, then, may lead to harms like being discriminated against or manipulated; such harm-centric framings are frequently used in court cases~\cite{angel2024distinguishingPrivacy, calo2014privacyHarmExceptionalizm} (about privacy violations) and legal scholarship~\cite{soloveDataWhatDataDoes2024}. This conceptualization avoids many of the limitations of the other formulations: it is oblivious to whether the data are private or sensitive, or accessed directly versus inferred from other data. 
Crucially, this framing may better explain and predict privacy-related behaviors, as privacy risks largely depend upon how data subjects perceive the effects of privacy harms~\cite{bhatiaEmpiricalMeasurementPerceived-2018}. Finally, technical and legal protective mechanisms can focus on preventing specific harms rather than aiming to eliminate data disclosure, which is close to impossible in practice. 

We operationalize this framing of privacy in an online study that investigates: (1)~What harms do data subjects anticipate when artificial intelligence (AI) algorithms infer personal data and use it to make consequential decisions? and (2)~How do perceived harms vary across data types as well as data subjects' demographic factors? We focus on two decision contexts: whether a student will drop out of a course or a job candidate should be hired---both are being increasingly AI-driven in practice~\cite{hunkenschroer_is_2023, dropouts-moocs, oulad_dropout_performance_prediction, predictive-LA-survey, noauthor_pdf_2024,noauthor_predictive_nodate, kodiyan2019-overview}. In this study, we surveyed college or university students (N=400), who are also current or future job seekers, and thus the population likely to be impacted the most by such algorithms (i.e., the ``experiential experts'' in this context~\cite{youngInclusiveTechPolicy-2019}). Each participant indicated whether they (dis)agree that the use of personal data will cause privacy harms in either the education or the employment context (between-subject design). The study included six classes of personal data (demographics, personality, traits, and emotional state) and 14 types of harms (e.g., manipulation, bias, and stereotyping), selected through a comprehensive literature review (see~\S~\ref{sec:methodology}). 


Our data reveals interesting and nuanced patterns in participants' conceptualization of privacy harms. For example, when demographic or personality trait data is used in either domain, participants anticipated harms like being stereotyped, discriminated, or manipulated; but perception of such data as inherently private was much weaker---lending support for the harm-centric understanding of privacy. Surprisingly, participants indicated that traits like creativity and motivation would cause harms like inconsistent and biased decisions when used in the education domain, but not when used to make hiring decisions. Finally, we uncover complex variations of perceived harms across participants' demographics. Overall, female, older, and non-white participants expressed more concerns about different harms compared to male, younger, and white participants, respectively; however, the type of harms they were concerned about varied across data type and the context. These findings not only provide a clearer picture of privacy violations but also offer actionable preventive measures. We additionally demonstrate the robustness of our measurement approach through reliability statistics and factor analysis, and practical significance by applying the harm categories in real-life incidents from the AIID dataset~\footnote{\url{https://incidentdatabase.io}}. Together, they prove that a harm-centric conceptualization of privacy can be valuable in many contexts that usable privacy researchers study. In sum, this paper makes the following contributions:

\begin{enumerate}
    \item We conceptualize and operationalize a harm-centric conceptualization of privacy and evaluate this framing on two decision contexts that are of increasing relevance to society.
    \item We empirically demonstrate the reliability and consistency of measuring privacy through perceived harms. Thus, we extend the literature on privacy frameworks. 
    \item We provide empirical evidence for the benefits of harm-centric framing. For example, privacy as a shield from harm detects violations of privacy that would go undetected under other frameworks (see results and discussions) and could guide technical and policy interventions to minimize harms when a complete ban on data collection is infeasible or undesirable. Another crucial benefit is that we show the existence of heterogeneity in harm perceptions: participants with different demographic properties significantly varied in what harms they were concerned about and how much, even for the same data. Thus, this framework may facilitate equitable privacy (compared to prioritizing the preservation of norms that are dictated by the majority).

    \item We discuss how this framing can be used for privacy threat modeling, and usable privacy research to understand people's privacy mental models and related behaviors. 
\end{enumerate}

Our findings will, we believe, directly impact the resolution of privacy harms caused by AI-based decision-making in education and employment. We also hope our framework will guide future research in understanding privacy in other contexts. 

%% file: sections/2_literature.tex
\section{Background and literature review}
\label{sec:literature}
\subsection{Privacy frameworks}
\subsubsection{Privacy as concealment of (or control over) data}\label{sec:concealment}
The bulk of privacy research conceptualizes privacy as the concealment of data that is considered private and declares that privacy is violated when such data is leaked (see chapter 2 of~\cite{knijnenburg_modern_2022} for a review). Such a conceptualization has many limitations. The boundary of `private' and `non-private' data is arbitrary and malleable, and data considered non-private can be used to infer `private' data~\cite{soloveDataWhatDataDoes2024}. This framing has been used to investigate people's mental models and concerns about private data leaks, which has unveiled phenomena like the so-called \textit{privacy paradox}---when people express concerns about data leaks but behave in ways that leak private data---that was later debunked~\cite{soloveMythofPrivacyParadox2021, knowlesUnParadoxingPrivacyConsidering-2023}. The focus on protecting private data is partly to blame for this paradox. In this increasingly networked world, with much of the social and professional activities being digitally mediated, people, even those with the strongest privacy concerns, have no choice but to use services that require information disclosure~\cite{guberek_keeping_2018, palenUnpackingPrivacyNetworked-2003}. Opting out of digital tools use is tantamount to opting out of society~\cite{richardsTakingTrustSeriously-2015}, and thus nondisclosure of private data is not viable~\cite{hullSuccessfulFailureWhat-2015}.

A related conceptualization of privacy is control over personal data~\cite{knijnenburg_modern_2022}, which is widely used in privacy laws and scholarship~\cite{soloveKafkaAgeAIPrivacy2024}. Yet, exercising this control is difficult, if not impossible, due to the same reasons mentioned above. On top of that, the current practice of providing that control to data subjects is typically through notice and consent, which, in reality, exacerbates privacy violations by creating unrealistic cognitive burden, desensitizing people to such notices, or being manipulative~\cite{solove_murky_2023}. Even if people were able to meaningfully exercise control over their data, they could do it only when the data was still undisclosed; this framing falls apart when one party (e.g., a data broker) gets access to the data and the data subject cannot exercise any more control for future transfers or use of that data. Finally, data obtained through consent can be subsequently used in a way that harms data subjects~\cite{solove_murky_2023}. Thus, the current conceptualizations of privacy fail to provide a useful framework for capturing all problems caused by privacy violations~\cite{angel2024distinguishingPrivacy}.




\subsubsection{Privacy as \textit{Contextual Integrity}} Nissenbaum's theory of privacy as contextual integrity (CI), which has been steadily gaining popularity over the last two decades, views privacy as the flow of data according to established norms and expectations in a given context~\cite{nissenbaum2009privacy}. As Nissenbaum stated in her seminal work, ``[w]hat people care most about is not simply restricting the flow of information but ensuring that it flows appropriately.''~\cite{nissenbaum2009privacy}. This formulation side steps problems like arbitrary classifications of private and public data, and contextualizing data flows offers concrete evidence of what data subjects consider as privacy violations (or not) as opposed to more general and sometimes vague notion of privacy concerns (which may mean nothing without a context); CI has been used in numerous studies to understand privacy (both normative and descriptive senses) as well as predict people's behaviors~\cite{kumarRoadmapApplyingCI2024}. 

Yet, CI's one of the major limitations is, as Nissenbaum explained, that it only considers flows of semantically meaningful data (such as a person's age) and excludes finer grained data (such as online browsing behaviors) or information that can be inferred from other data (e.g., inferring someone's age from a photo posted online)~\cite{nissenbaumContextualIntegrityDataChain2019}. Thus, CI cannot explain privacy violations in a large amount of the data flows in today's digital world, which is driven by artificial intelligence (AI) and machine learning-based tools that predict or infer information based on user data. 

Another, more fundamental limitation of CI is that, since it relies on norms to define appropriate data flows, it cannot demonstrate privacy violations resulting from established behaviors. For example, if posting photos and videos online has become ``normal'' in a society, then CI cannot identify any privacy violations from such practices, even as technological advancements create new privacy harms (such as using recently developed generative models to create deep fake pornography from previously posted videos~\cite{umbachDeepFakePornography2024}). As Van de Poel argues, ``the inherent conservatism of CI making it possible to justify problematic technologies that follow established norms''~\cite{vandepoelSociallyDisruptiveTechnologies-2022}.


\subsubsection{Harm-centered framing of privacy}
Instead of asking 'what data needs to be protected' or 'which data flows need to be prevented' to protect privacy, we ask 'what harms can result from the use of personal data' and conceptualize privacy as a shield against those harms. 
This formulation is based on various theoretical and applied research in privacy, law, and social science. The realization that Privacy is much more than data protection~\cite{angel2024distinguishingPrivacy}, privacy's intimate relationships with many other human values---such as freedom, protection from manipulation, and identity building~\cite{richardsWhyPrivacyMatters-2021}---as well as privacy's role in preventing harms, such as discrimination and algorithmic manipulation, have long been recognized in social and legal scholarship. Increasingly, researchers have begun to identify different types of information-based harms as privacy harms~\cite{angel2024distinguishingPrivacy}. In his recent work, prominent privacy scholar Daniel Solove made a strong case for focusing on privacy harms from data use, rather than on the data itself~\cite{soloveDataWhatDataDoes2024}. Solove gives the example of a religious leader who might very much want people to know his role (for maximum impact) but would not want this information to be used to discriminate against him. Effective privacy protection would mitigate this negative effect, rather than keeping the information secret. As Solove argued elsewhere~\cite{solove2015meaningValue}: ``A theory of privacy should focus on the problems that create a desire for privacy.''

A harm-centric conceptualization of privacy sidesteps the issues with defining what data are private or is not concerned about how data gets leaked. Rather, it makes a case against using data in ways that harm data subjects, even if consent was obtained or the data is already public. As opposed to norms, which by definition are dominated by the majority, the harm-centric framing allows uncovering and addressing negative consequences faced by population groups of any size and composition. Importantly, this framing can potentially explain and predict people's behaviors better than other frameworks, since harms are the realization of risks, and past research has shown that concrete risk understanding aids decision-making better than abstract privacy concerns~\cite{investigate-risk-perception}. Additionally, past studies reported that people are concerned about how data can be used in harmful ways rather than about the act of merely providing the data~\cite{jakobiItWhatTheyDoWithData2019}, and that people behave in ways to avoid negative consequences from technology use, even if they do not actively think about privacy~\cite{colnagoThereReversePrivacy-2023}. Thus, harm-centric framing can be useful in encouraging privacy-conscious behaviors, which has been a major goal in usable security and privacy research. Finally, focusing on privacy harms can provide clarity about the nature of privacy protections afforded by computational mechanisms. For example, differential privacy, one of the most popular techniques, provides a quantifiable privacy guarantee. Thus, if an AI model is trained in a differentially private manner, any information extracted from that model cannot be linked back to individuals whose data was used in the training. Such a model, however, still may harm other people's privacy; e.g., after deployment, an AI-based hiring agent learning secrets about job candidates. These issues will be characterized as privacy harms under the proposed framework. 

Despite compelling benefits, measuring privacy in terms of harms in a given context has not been formally explored in the literature. Several past works, however, empirically identified privacy harms. For example, Karwatzki~\etal identified and empirically validated seven types of harms (e.g., physical, social, and psychological) that individuals perceive if their privacy is invaded~\cite{karwatzkiYESFIRMSHAVE-2018}. Jacabi~\etal expanded this research by eliciting 91 privacy harms  (e.g., privacy intrusion, discrimination, and behavioral manipulation) through focus group interviews and workshops. Later, they employed experts to categorize those harms to create a high-level taxonomy~\cite{jakobiTaxonomyUserperceivedPrivacy-2022a}. Solove's seminal work created a taxonomy of privacy~\cite{soloveTaxonomyPrivacy} that includes 16 activities (e.g., surveillance) that lead to privacy harms, whereas our aim is to surface privacy harms from a given activity (e.g., AI using inferred personal data).

Outside academia, government agencies have started to adapt risk-based (the product of the impact of a privacy harm and its likelihood) frameworks to privacy, often conceptualized as the negative effects as perceived by data subjects~\cite{bhatiaEmpiricalMeasurementPerceived-2018}. For example, NIST (National Institute of Standards and Technology) has begun to frame privacy issues in terms of risk, drawing parallels to security violations as loss of confidentiality, integrity, and availability of a system~\cite{brooksIntroductionPrivacyEngineering-2017}. 

These works demonstrate the necessity and practical usefulness of focusing on privacy harms; however, a formal treatment to \textit{measure privacy as a shield against specific harms in a given context} has yet to be done. Such a formalization is essential to make theoretical advancement in understanding privacy itself and complementing existing privacy frameworks.

\subsubsection{Procedural justice and privacy harms}
We use the lens of procedural justice theory~\cite{leventhal1980} to formalize harm-centric privacy. Procedural justice examines if the process of a decision-making task is just and fair toward people who might be impacted by that decision~\cite{leventhal1980}. Such investigations are empirical and bottom-up, thus, they require surfacing the perceptions of the affected population (e.g., see~\cite{gilliland_perceived_1993, morse_ends_2022, grgic-hlaca_human_2018}. To mitigate harms from algorithmic predictions, legal scholars Crawford and Schultz proposed the concept of `the right to procedural data due process'~\cite{crawford2014bigDataDueProcess}---that is, individuals should be able to demand that predictive models be built following a just process to minimize negative impacts on them. This paper operationalizes this idea in the contexts of automated decision-making in education and employment through an empirical study.

\subsection{Use of AI in education and employment}\label{sec:rw:ai}
We briefly review the literature on AI use and associated privacy concerns in education and employment, two contexts this study focuses on.

\subsubsection{AI-based prediction tasks}
 AI and machine learning-based models have become ubiquitous in both sectors. Some of the most common predictive tasks in education include course performance prediction~\cite{predicting_performance_waheed_2020} the probability of dropping out from a course~\cite{predictive-LA-survey}, students' attention and behavior prediction~\cite{kaurSystematicReviewPrediction-2020}, as well as detecting students' cheating behaviors during remote exams~\cite{tiongEcheatingPreventionMeasures-2021}. In employment, AI models evaluate potential candidates’ job performance~\cite{noauthor_predictive_nodate}, likelihood of a candidate accepting a job offer\cite{gaba_predictive_2024}, and predicting employee turnover and retention\cite{punnoose_prediction_2016}. Additionally, AI systems are being employed to screen resumes, conduct initial interviews, and even make hiring recommendations~\cite{lacroux_should_2022}. 

\subsubsection{Use of personal data and resulting privacy harms}
Various types and modalities of personal data have been used to train predictive models in education, including demographics and socio-economic status, behavioral data (login patterns, time spent on tasks, interaction with resources), and even mobility and activity patterns. Likewise, AI in employment uses resume and online portfolio, assessment scores (cognitive ability tests, job knowledge tests), interview data (verbal responses, facial expressions, voice patterns), and external information (social media profiles, public records)~\cite{cassidy2024, Alene2022}.

Privacy researchers and activists have raised numerous concerns about AI models using such data~\cite{edtech-pets22}, and many laws prohibit using many of these data items as decision criteria (e.g., gender and race for college admission or hiring employees). However, recent research has shown that legal measures to prevent direct use of these data items do not eliminate privacy harms. For example, many behavioral data encode demographics and socio-economic information, as well as physiological and psychological characteristics~\cite{edtech-pets22, overlearning, hendersonSelfDestructingModelsIncreasing-2023}; these information can be inferred by re-purposing AI models (originally trained to predict, e.g., students' course performance) or extracted from those models' internal representation of input data~\cite{edtech-pets22, hendersonSelfDestructingModelsIncreasing-2023}. Moreover, even in the absence of an active adversary, the models may implicitly rely on such data to make decisions, resulting in biased, exclusionary, and stereotyped decisions. Concerningly, many AI systems deployed in the real world are already exhibiting such behaviors~\cite{wilsonGenderRaceIntersectional-2024, obermeyerDissectingRacialBiasHealth2019, kodiyan2019-overview}. 

Unfortunately, the current conceptualizations of privacy fall short in characterizing such incidents as `violations of privacy'~\cite{angel2024distinguishingPrivacy}, as data is already out of control (public) or voluntarily provided by the data subjects. However, the role of privacy, as broadly understood, includes providing freedom from manipulation, discrimination, and marginalization, among other things; and such framing would rightly categorize these incidents and privacy harms, of which this paper provides empirical evidence.

%% file: sections/3_methodology.tex
\section{Methodology}
\label{sec:methodology}
\subsection{Study design}

\subsubsection{Domain and target population selection}

An investigation of harm-centric conceptualization of privacy needs to be domain or context-dependent, since using the same personal data in different domains can result in different harms. We select two domains: education (AI predicting the probability of students dropping out of courses) and employment (AI recommending job candidates). We select them, as \S~\ref{sec:rw:ai} details, AI has become ubiquitous in both of these contexts, and importantly, they can raise privacy issues without any disclosure of `private data' and do not require an active adversary.


To be comprehensive and inclusive, such investigations must also be bottom-up: gathering empirical evidence of harms perceived by people whose personal data will be used for these decision-making~\cite{magassaInclusiveJusticeApplying-2024, youngInclusiveTechPolicy-2019, jakobiTaxonomyUserperceivedPrivacy-2022a}. Ideally, they are also the \textit{experiential experts}: people living the experience or those closely associated with someone living the experience~\cite{youngInclusiveTechPolicy-2019}. The legal domain often identifies privacy harms deductively (by basing on fundamental human rights violations), where scholars have been recently raising critical questions about authority, legitimacy, and inclusiveness when identifying new privacy harms~\cite{angel2024distinguishingPrivacy}. To avoid such pitfalls, we gather data from college and university students, who are also current or future job seekers, and thus likely have experience with AI-based decision-making in those contexts. 

With this road map, we designed and conducted an online study to understand how students perceive the harms of using personal data by AI models to make decisions in education and employment. 

\subsubsection{Selection of data types}

We focus on six types of personal data as shown below; their selection was guided by three criteria: (1)~data commonly used by AI systems in education and employment contexts, (2)~data that are expected to generate a wide range of perceived harm level (i.e., high variance), and (3)~data that may not be directly collected but can be inferred from other data.

\begin{itemize}
\item \textbf{Demographics}: The use of demographic information, such as age, gender, and socio-economic variables, as predictors in AI models has been extensively studied in the context of privacy, ethics, and fairness of algorithmic systems~\cite{edm_and_privacy, using_demographics_in_edm_2020}. While in many cases their direct inclusion as a decision criterion is prohibited by law, past research has shown that such information may be encoded in other predictors (such as behavioral data~\cite{edtech-pets22, overlearning}) and thus might indirectly induce discriminatory behaviors of AI models.

\item \textbf{Personality traits}: Personality traits, such as extraversion and agreeableness~\cite{big-five-original}, have been extensively studied as predictors of academic performance~\cite{hakimiRelationshipsPersonalityTraits-2011} and employability~\cite{timmons_pre-employment_2020}. Like demographics, personality traits are also linked to other data such as resumes and online profile information~\cite{kamble_innovative_2022, van_mil_promises_2021}, as well as social media activity and behaviors during interviews~\cite{kamble_innovative_2022, giritlioglu_multimodal_2021}, and can indirectly influence decisions of AI models, raising ethical and fairness concerns. 

\item \textbf{Emotional states}: As AI tools continue to improve in inferring emotional states from multi-modal data, such as facial expressions and body movements during remote video interviews, they raise concerns about privacy, accuracy, and potential misinterpretation of emotional data raise ethical challenges, as such technologies may violate emotional privacy and perpetuate biases or discrimination~\cite{roemmich_emotion_2023,roemmich_values_2023}.

\item \textbf{Motivation:} Student motivation is frequently used as a predictor in both educational and employment contexts\cite{ligeiro2024recruitment}. Motivation is generally considered volitional and within one's control, but concerns arise regarding privacy, accuracy of measurement, and potential for manipulation when used as a decision criterion by AI systems.

\item \textbf{Creativity and Problem Solving:} Creativity and problem-solving abilities are increasingly valued in both educational assessment and hiring decisions\cite{marrone2022creativity}. These attributes can be measured through standardized tests, portfolio assessments, situational judgment tests, and analysis of past creative outputs. While these abilities are considered trainable to some extent, their measurement often involves subjective evaluation, raising concerns about consistency, reliability, and potential bias in AI-based assessment.

\item \textbf{Physical or Cognitive Impairment:} Information about physical or cognitive impairments may be directly disclosed by individuals seeking accommodations or indirectly inferred through behavioral patterns, interaction data, or performance metrics \cite{nugent2020recruitment}. The use of such information in decision-making raises significant ethical concerns, particularly regarding discrimination and compliance with disability rights legislation\cite{buyl2022tackling}.

\end{itemize}

\subsubsection{Selection of privacy harms}


We identified 14 harms based on the literature on human-centric privacy, law, ethics, procedural justice and fairness, and recent works on social harms from algorithmic decision-making. This set includes both material (objective) and psychological (subjective) harms~\cite{caloBoundariesPrivacyHarm-2011}. Table~\ref{tab:harm-types} provides their descriptions and references to the literature where they were studied in the context of data privacy. 


\begin{table*}[]
    \centering
    \footnotesize
    \begin{tabularx}{0.99\textwidth}{l X}
    \toprule
    \textbf{Harm} & \textbf{Definition}\\
    \midrule
    Invasion of privacy & Data is considered inherently private, and just their unwarranted exposure can lead to discomfort and chilling effects~\cite{abercrombie_collaborative_2024, jakobiTaxonomyUserperceivedPrivacy-2022a}.\\
    
    Bias & Discriminatory outcomes due to the use of certain data for decision-making~\cite{gdpr, barocas2016bigdata, leventhal1980}. \\
    
    Wrong inference & The predictor (e.g., age) may not itself be accurately inferred from raw data~\cite{grgic-hlaca_human_2018, wachter2019reasonable}\\ 
    

    Loss of autonomy & AI can hamper individuals' ability to make informed, voluntary decisions about their lives or actions ~\cite{ahmad2023privacy, susser2019manipulation}. \\
    
    Stereotyping & Leaked information can be used for reductive generalizations about individuals based on race, gender, or other social categories~\cite{caliskan2017semantics, abercrombie_collaborative_2024}. \\
    
    Violation of ethics & Subjecting individuals to algorithmic predictions solely for the service provider's benefits and efficiency can be unethical and an arbitrary exertion of power~\cite{crawford2014bigDataDueProcess, akgun2021ethics, martinPredatoryPredictionsEthics-2023} \\
    
    Inconsistency & Relying on certain predictors can lead to inconsistent decisions~\cite{leventhal1980, hanson2021ofqual} \\
    
    Lack of reliability & Certain data may not be a reliable (e.g., non-causal) predictor~\cite{grgic-hlaca_human_2018, gdpr} \\
    
    Defamation & Inferred data can lead to defamation if leaked~\cite{abercrombie_collaborative_2024, tenzer2023defamation}. \\
    
    Harassment & Inferred data can be used for online/in-person  harassment~\cite{brundage2018malicious, abercrombie_collaborative_2024}. \\
    
    Manipulation & Personal data can be used to manipulate (e.g., people's beliefs or behaviors)~\cite{susser2019manipulation, abercrombie_collaborative_2024}. \\
    
    Stress & Data subjects can suffer from stress or anxiety if personal data about them is used~\cite{apa2023stress}. \\
    
    Inaccurate decision & The use of certain predictors can lead to inaccurate decisions~\cite{leventhal1980, wachter2019reasonable}. \\
    
    Lack of user control & Certain predictors may be perceived as unjust to use since an individual may have no control over it (e.g., gender and ethnicity)~\cite{leventhal1980, grgic-hlaca_human_2018} \\
    
    \bottomrule
    \end{tabularx}
    
    \caption{Harm types and their descriptions}
    \label{tab:harm-types}
\end{table*}


\subsection{Data collection}

After obtaining approval from our institutional ethics board, we implemented this study on Qualtrics. We recruited participants on Prolific~\cite{prolific}; we indicated that our target population consists of current college students (including those at community colleges) and university students. Each participant answered questions related to either the education or the employment context (randomly selected).

In the survey, after consenting, participants were briefed about the use of AI in education (or employment, depending on the randomly assigned scenario). Following that, we explained how an AI model can infer or predict an attribute (e.g., demographics) using various data sources (e.g., images found online) and use that in predicting dropout risk (or fit for a job). After that, we asked them to indicate their (dis)agreement with statements about such inferences being harmful (e.g., ``Use of personality traits to decide dropout risk can lead to biased decisions'') using a five-point Likert scale (``Strongly Disagree'' to ``Strongly Agree,'' coded from -2 to 2 during analysis). To ensure data quality, we included one question hidden from human eyes but visible to an automated survey parser algorithm. Additionally, we included one qualitative question at the end and manually reviewed the answers to identify gibberish responses. The instructions and full questionnaire are provided in the Appendix (\S~\ref{sec:ques}). We first recruited 20 participants and used their responses to conduct a power analysis, which indicated that we need at least 392 participants for a significance test at a false positive level of 0.05. Next, we collected data from 430 more participants, but discarded 30 responses because of the high likelihood of AI-generated responses (because they answered the hidden question, N=4) or inattentive responses (detected by manual review of qualitative responses, N=26). The median time for study completion was 16 minutes and 40 seconds. All participants (except the 4 AI bots) were compensated with \$5 for completing the study.

\paragraph{Limitations.} Our study may suffer from sample bias since we rely on an online platform to collect data. Additional study may be needed to investigate if our results generalize to the whole US population of college and university students. We limited our investigation to six attributes (data types) to keep the study length reasonable. Another important limitation is that we investigated privacy harms in contexts that required the use of personal data. However, mere surveillance can cause privacy harms (e.g., creating chilling effects and inhibiting the freedom of expression~\cite{soloveTaxonomyPrivacy} even if no data is collected or used. We leave such investigations to future research.

%% file: sections/4_results.tex
\section{Results}
\label{sec:results}

\subsection{Participants}

Among the 400 respondents, 45.3\% (N = 173) self-identified as males, 52.1\% (N = 199) self-identified as females, 2.4\% (N = 9) self-identified as non-binary/third gender, and the rest preferred to self-describe (N = 1). Participants' age distribution showed the following: between 18-24(42.2\%, N = 161), between 25-34 (38.5\%, N = 147 ), between 35-44 (13.6\%, N = 52), between 45-54 (4.7\%, N = 18), and 55-64 (N = 4). Participants racial distribution showed the following: White (58.2\%, N = 219), Black or African American (28.5\%, N = 107), Asian (9.3\%, N = 35), American Indian or Alaska Native (2.7\%, N = 10), and Native Hawaiian or Other Pacific Islander (N = 5). A majority of participants indicated that they were Undergraduate/Bachelor (53.3\%, N = 201), followed by Masters (19.4\%, N = 73), Technical/Community College (16.5\%, N = 62), Doctoral (9\%, N = 34), and others (N = 7). Participants were classified based on their field of study: STEM (47\%, N=189), and Non-STEM(53\%, N=211).  

\subsection{Measurement reliability}
We computed Cronbach's alpha, which indicates internal consistency of a set of items~\cite{tavakol_making_2011}, in this case, the 14 harm statements. Table~\ref{tab:education_alpha_itc} shows the alpha values for the education context (the first row), the minimum alpha being 0.93, the items indicate high reliability~\cite{tavakol_making_2011}. The subsequent rows in Table~\ref{tab:education_alpha_itc} show alpha scores when one item was removed at a time. Removing any item decreased the score, indicating their complementary roles. Each item also correlated highly with the mean of the other items (i.e., item-total correlation, shown within parentheses). We obtained similar results for the employment context, but left them out due to space constraints. 

\input{sections/cronbach_alpha_table}

\begin{figure}[h]
    \centering
    \includegraphics[width=.48\textwidth]{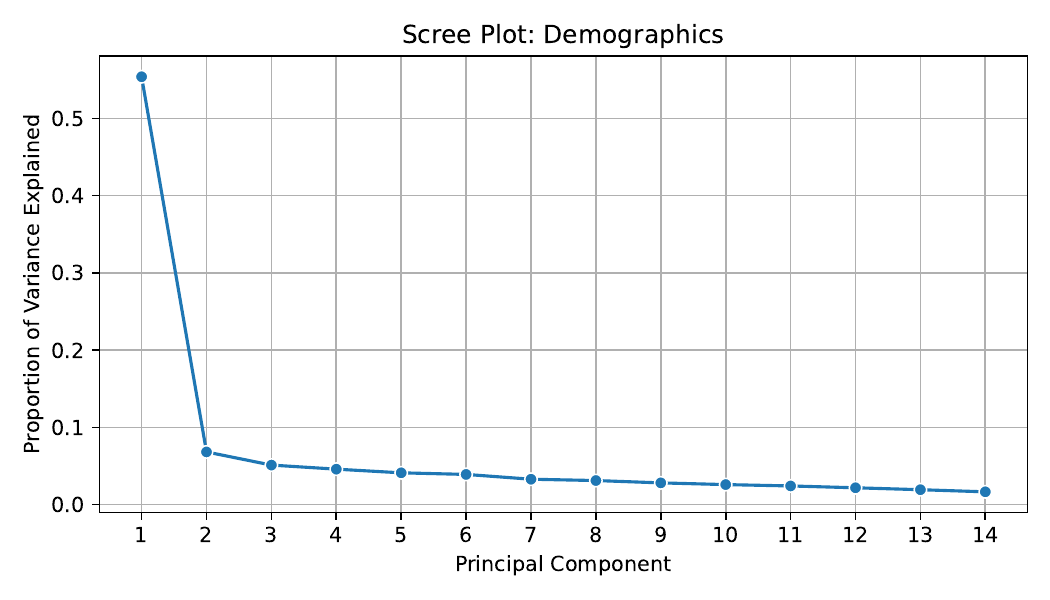}
    \caption{Scree plot for factor analysis}
    \label{fig:demographics_scree_plot}
\end{figure}

Next, we conducted factor analysis to assess the dimensionality of the latent variable measured by the 14 items. We varied the number of dimensions from $1$ to $14$, created scree plots, and used the elbow method to select the final number of dimensions~\cite{field2012discoveringStatR}. The scree plot for Demographics data type in education context is shown in Figure~\ref{fig:demographics_scree_plot}, where the explained variance sharply declines after the first factor, indicating a single latent dimension. All of the remaining 11 plots showed the same pattern, we omit them due to space limitations. The total variance explained ranged from 52\% to 66\% in the education context and 50\% to 61\% in the employment context. These results indicate that the single factor sufficiently contains information about the items (50\%--60\% variance is considered a good fit in social science research~\cite{field2012discoveringStatR}. In sum, these results indicate that all items are measuring one underlying construct: \textit{privacy harm}. Thus, the sum across all items would indicate a measure for the ``total privacy harm'' from using a specific personal data.

\begin{SummaryBox}{Takeaways}
    Statistical analysis demonstrated that all 14 items are internally consistent and reliably measure a single latent construct: \textbf{privacy harm}. Thus, this approach can be used in other domains to measure privacy through perceived harms.
\end{SummaryBox}


\subsection{Perceived harms from private data use by AI Models}
\begin{figure*}[htbp]
\centering
    \begin{subfigure}[b]{0.48\textwidth}
        \includegraphics[width=\textwidth]
{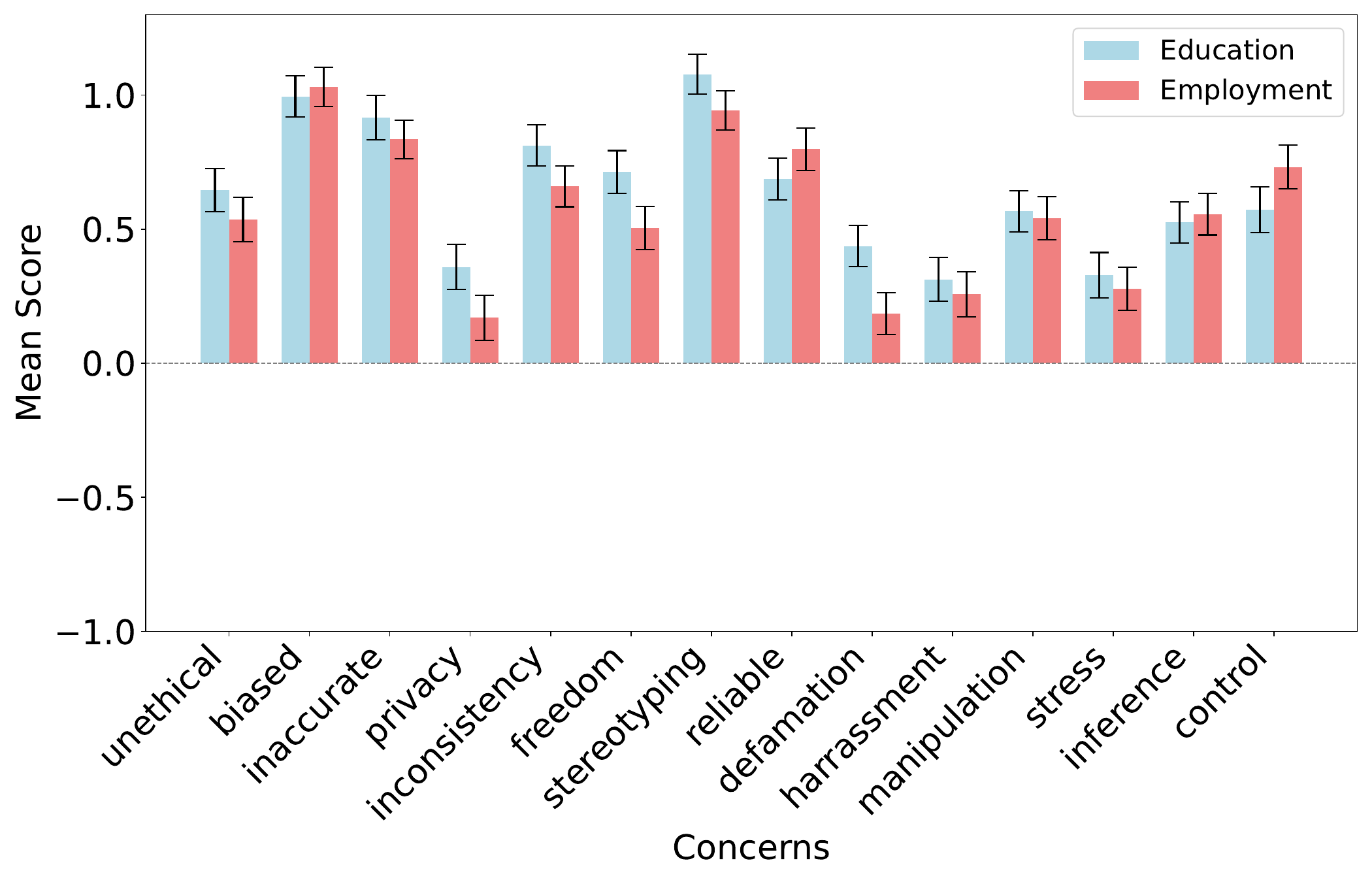}
        \caption{Demographics}
        \label{fig:demographics}
    \end{subfigure}
    \hfill
    \begin{subfigure}[b]{0.48\textwidth}
        \includegraphics[width=\textwidth]{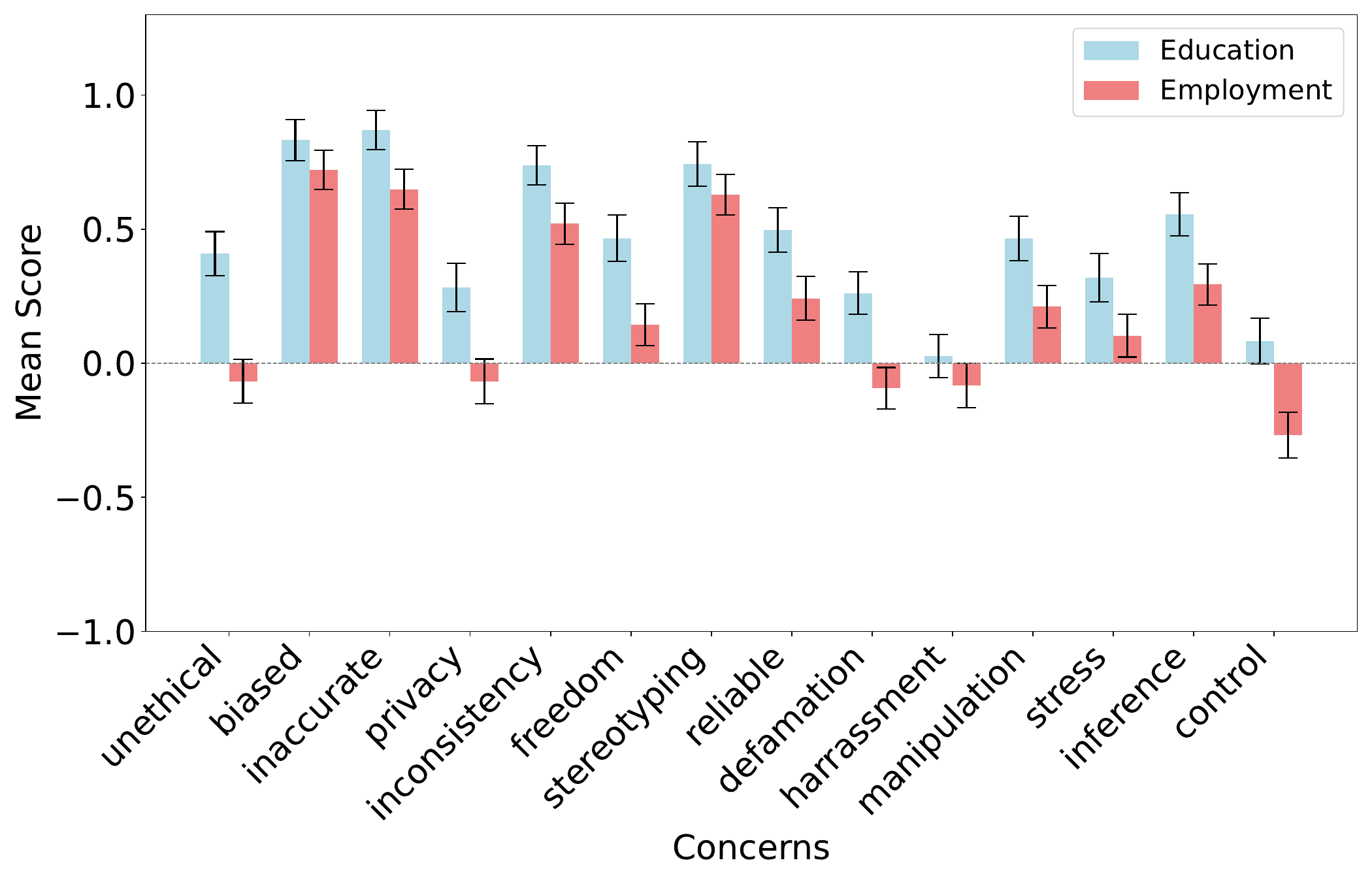}
        \caption{Personality Traits}
        \label{fig:personality_traits}
    \end{subfigure}
    
    \vspace{0.5cm}
    
    \begin{subfigure}[b]{0.48\textwidth}
        \includegraphics[width=\textwidth]{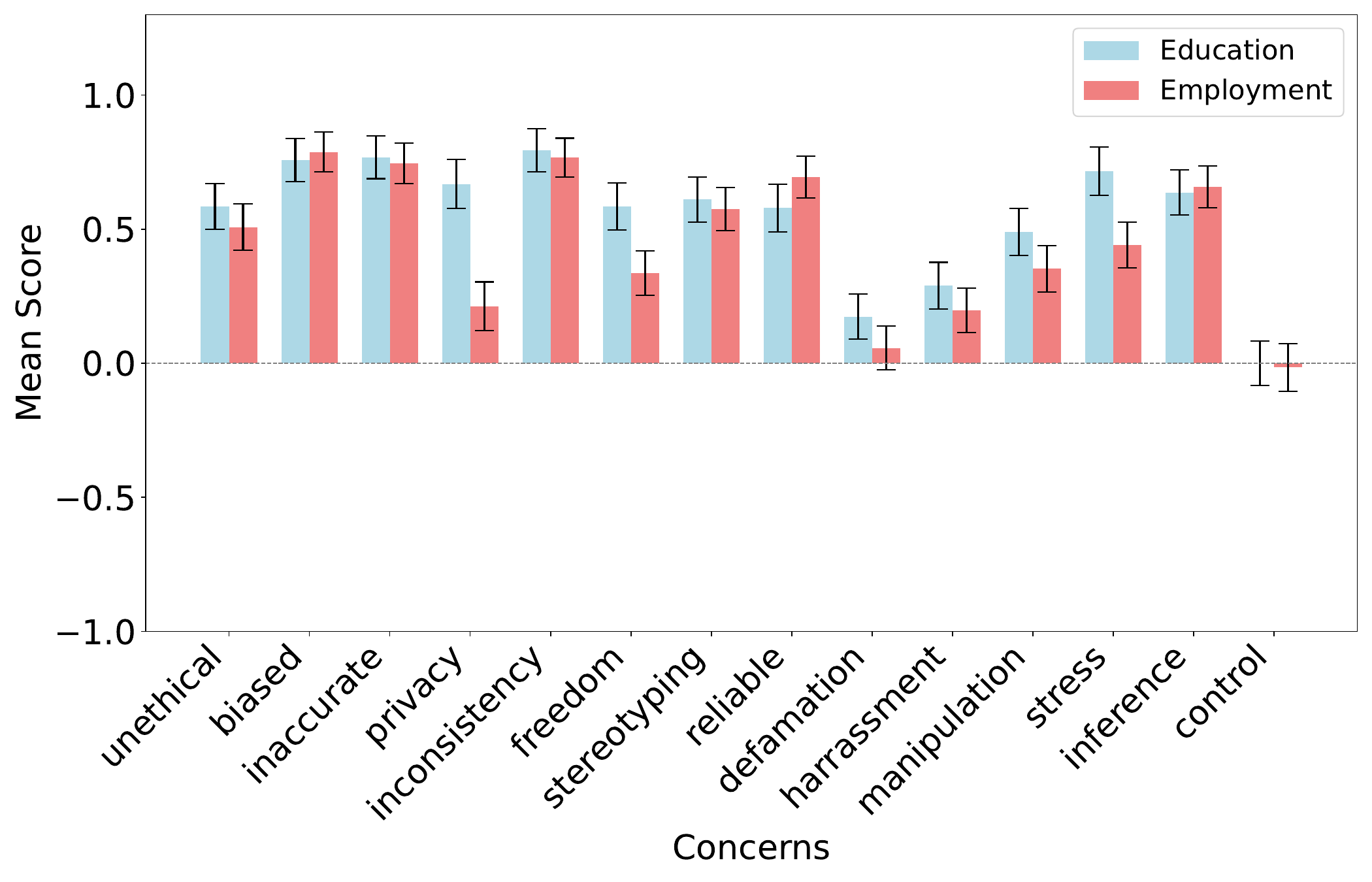}
        \caption{Emotional State}
        \label{fig:emotional_state}
    \end{subfigure}
    \hfill
    \begin{subfigure}[b]{0.48\textwidth}
        \includegraphics[width=\textwidth]{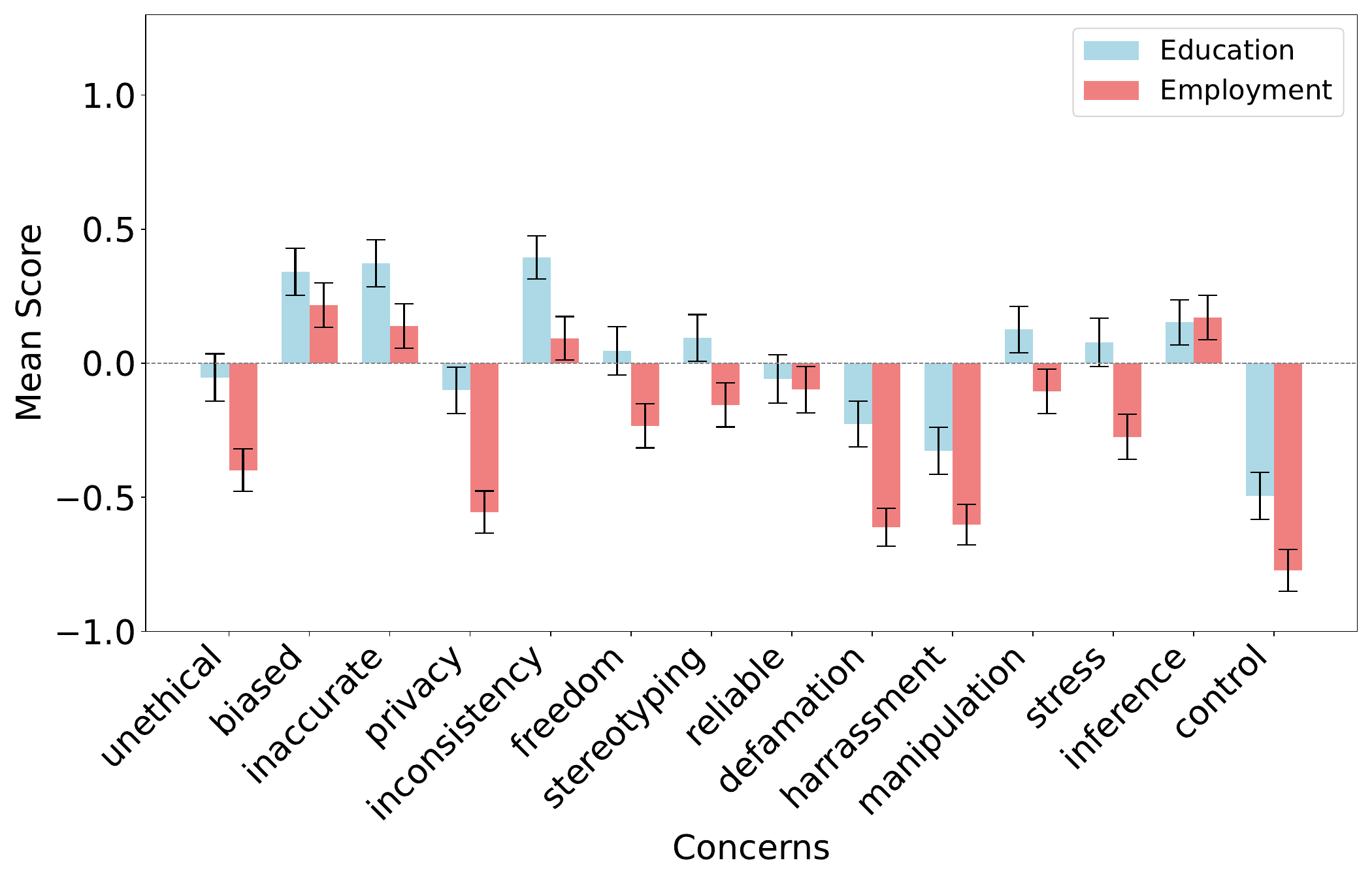}
        \caption{Motivation}
        \label{fig:motivation}
    \end{subfigure}
    
    
    \begin{subfigure}[b]{0.48\textwidth}
        \includegraphics[width=\textwidth]{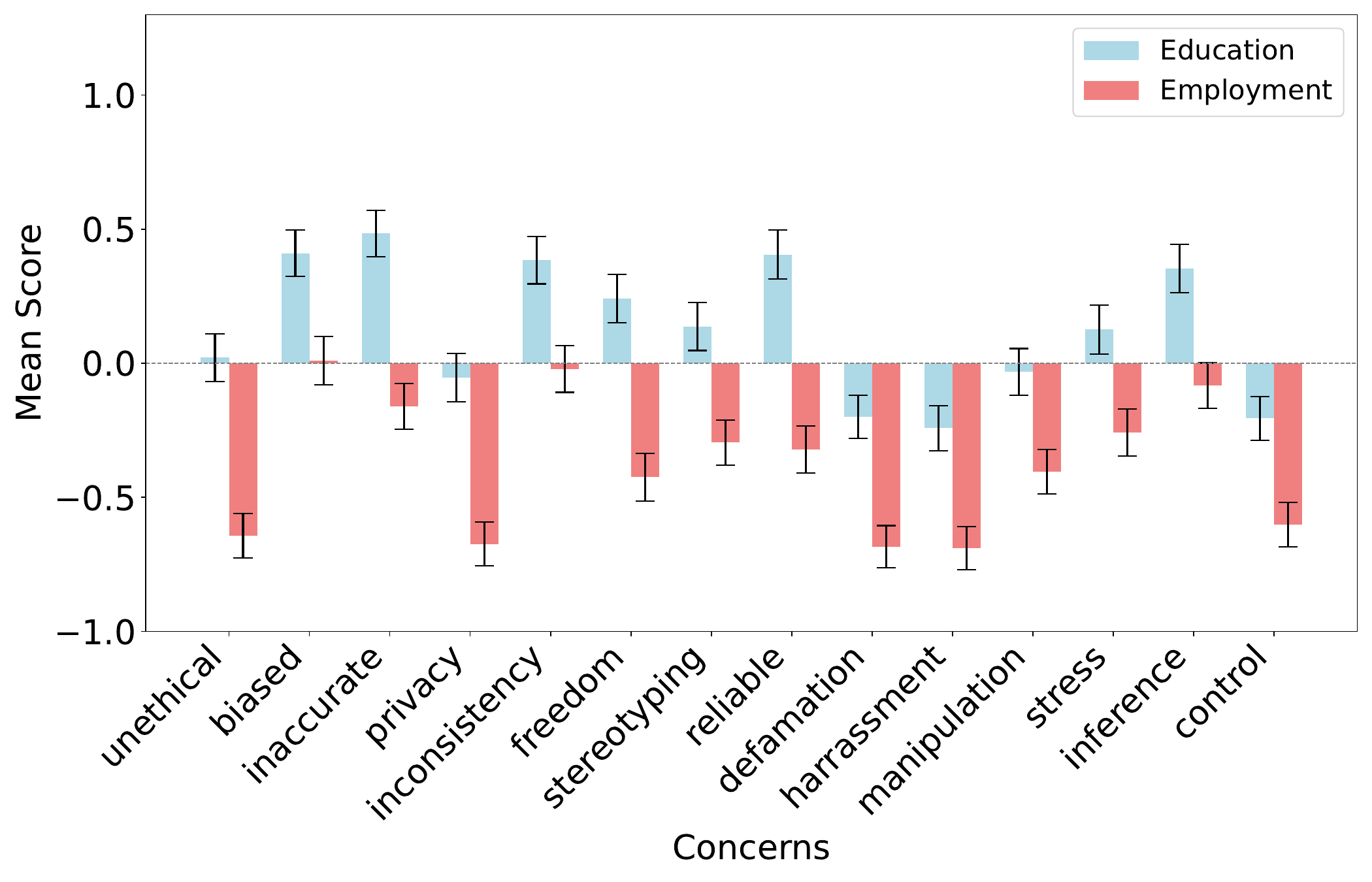}
        \caption{Creativity and Problem Solving}
        \label{fig:creativity_and_problem_solving}
    \end{subfigure}
    \hfill
    \begin{subfigure}[b]{0.48\textwidth}
        \includegraphics[width=\textwidth]{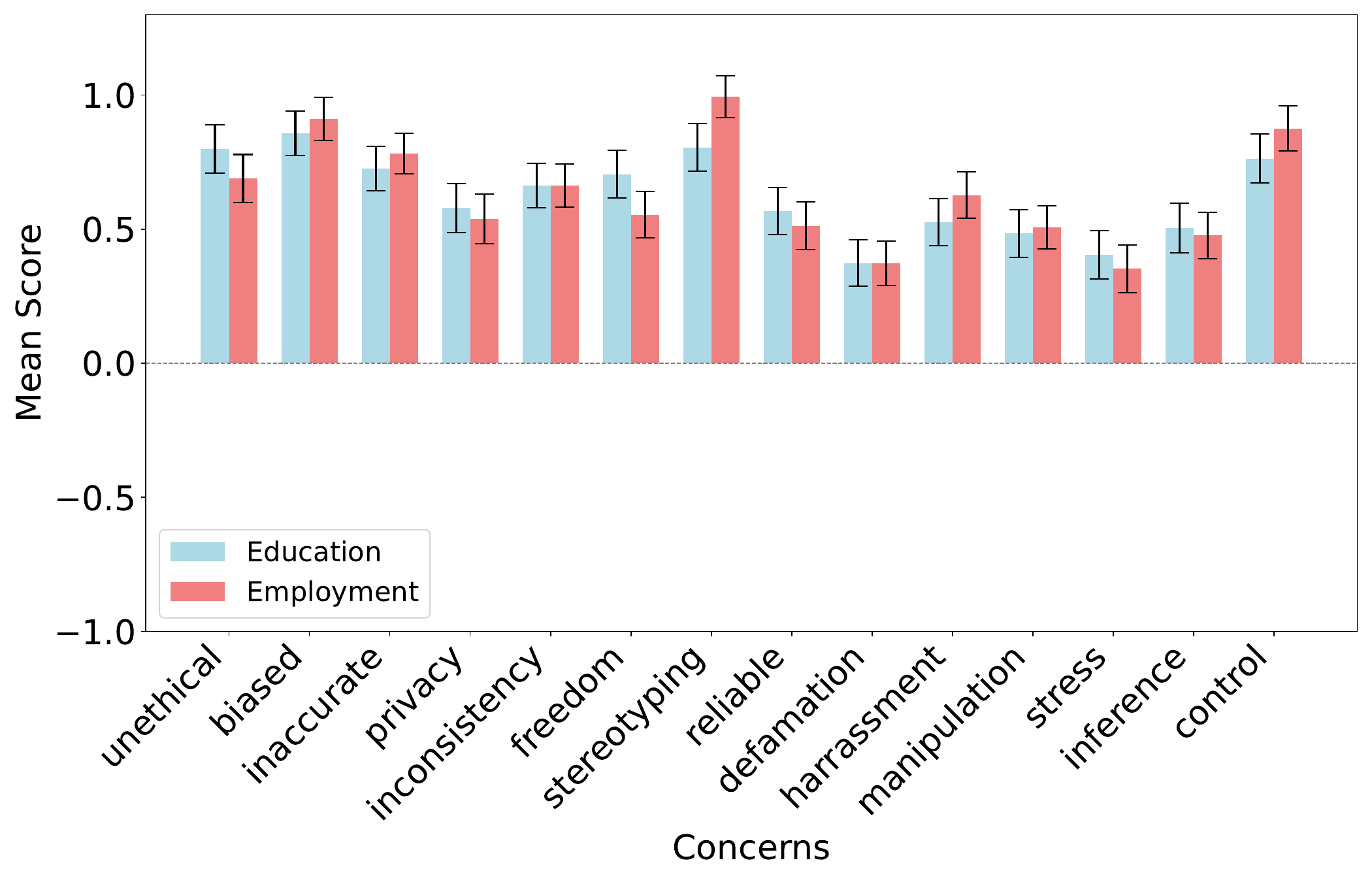}
        \caption{Physical or Cognitive Impairment}
        \label{fig:physical_or_cognitive_impairment}
    \end{subfigure}

    \caption{Mean and SD of perceived harms in education and employment. }
    \label{fig:all_attributes_comparison}
\end{figure*}

This section presents which harms that participants perceive to be likely in the context of personal data use for dropout prediction and hiring job candidates.  Figure~\ref{fig:all_attributes_comparison} shows the mean and standard error of participants' ratings along the 14 harm dimensions for the six data types. 

Overall, participants agreed that almost all harms can materialize when information about demographics, emotion, or disability status is learned and used by AI models. The level of agreement was also comparable across most harm categories between education and employment contexts. Using personality traits was associated with bias, inaccurate prediction, and stereotyping in both contexts, but other harms were perceived only in the education context. Results between these two contexts diverged even more for the other three data types: using problem-solving skills or work motivation was perceived to be harmful in the education context but not in the employment context. Interestingly, participants mostly disagreed that these personal data are inherently private (except for disabilities and emotional states), especially in the employment context. These results highlight that data considered non-private can lead to privacy harms, and the same data can lead to different harms in different contexts. 

Beyond general trends, separate plots for education and employment contexts, showing the numbers of participants in each of the response categories (see Figure~\ref{fig:response_counts} in the Appendix). They show how prevalent each of the perceived harms is among the participants. Importantly, they reveal the non-uniformity of harm perceptions among groups of participants, and that at least some people feel vulnerable to each of the harms, even if the general trend is negative (e.g., \textit{Defamation} resulting from \textit{Motivation}). Such a feeling of vulnerability may correlate with people's demographic and socio-economic factors, allowing us to identify the group(s) requiring the most protection against a certain harm.

\begin{SummaryBox}{Takeaways}
\begin{enumerate}
    \item The use of demographic, emotional state, or physical/cognitive disability status in either context was perceived as most concerning. 
    \item Use of data not inherently private can cause privacy harms.
    \item For the same data, perceived harms can differ across contexts. Additionally, this approach surfaces harms that may impact only a minority of the population.
\end{enumerate}
\end{SummaryBox}

\subsection{Variations in harm perceptions}
We investigated whether concern levels for the same attribute vary across the gender, age, race, and education levels of the participants. Specifically, we compared males with females, participants 18--24 years old with those older than 24 years, `white' with  `non-white' participants, and undergraduate students with post-graduate students (resulting group sizes are shown in Table~\ref{tab:categroical_comparison}). The following sections detail the comparison results. 

\subsubsection{Education context}
We omit reporting all statistical tests due to space limits; rather, we present the comparisons that were statistically significant across participants' demographics (after applying Bonferroni correction for multiple tests).


\paragraph{Gender.} 
Female-identifying participants expressed a higher level of concern about many harms when personality traits or emotional state are used, compared to male-identifying participants; the reverse was not true in any case. 


For personality trait prediction, females expressed more concerns about prediction accuracy ($M_m=0.58$, $SD_m=1.10$, $M_f=1.10$, $SD_f=0.91$, $p<0.001$), ethics($M_m=0.19$, $SD_m=1.17$, $M_f=0.58$, $SD_f=1.15$, $p=0.02$), fairness($M_m=0.60$, $SD_m=1.15$, $M_f=1.03$, $SD_f=0.99$, $p=0.01$), privacy($M_m=0.06$, $SD_m=1.28$, $M_f=0.45$, $SD_f=1.27$, $p=0.03$), consistency($M_m=0.53$, $SD_m=1.10$, $M_f=0.91$, $SD_f=0.94$, $p=0.01$), autonomy($M_m=0.21$, $SD_m=1.27$, $M_f=0.69$, $SD_f=1.15$, $p=0.01$), stereotyping($M_m=0.40$, $SD_m=1.27$,  $M_f=1.03$, $SD_f=1.00$, $p= 0.00$), reliability($M_m=0.21$, $SD_m=1.12$, $Mean_f=0.70$, $SD_f=1.16$, $p=0.00$), defamation ($M_m=0.00$, $SD_m=1.15$, $M_f=0.47$, $SD_f=1.03$, $p=0.003$), manipulation($M_m=0.27$, $SD_m=1.22$, $M_f=0.62$,$SD_f=1.11$, $p=0.04$), stress($M_m=0.04$, $SD_m=1.27$, $M_f=0.54$, $SD_f=1.25$, $p=0.01$), wrong inference($M_m=0.28$, $SD_m=1.14$, $M_f=0.77$, $SD_f=1.08$, $p=0.00$). 

For emotional state, females were more concerned about ethics($M_m=0.36$, $SD_m=1.18$, $M_f=0.75$, $SD_f=1.20$, $p=0.03$), inaccuracy ($M_m=0.54$, $SD_m=1.11$, $M_f=0.95$, $SD_f=1.12$, $p=0.01$), privacy ($M_m=0.28$, $SD_m=1.33$, $M_f=1.00$, $SD_f=1.15$, $p=0.00$), consistency ($M_m=0.61$, $SD_m=1.14$, $M_f=0.95$, $SD_f=1.11$, $p=0.04$), autonomy ($M_m=0.33$, $SD_m=1.29$, $M_f=0.81$, $SD_f=1.16$, $p=0.01$), stereotyping ($M_m=0.36$, $SD_m=1.29$, $M_f=0.81$, $SD_f=1.09$, $p=0.01$), reliability ($M_m=0.31$, $SD_m=1.22$, $M_f=0.78$, $SD_f=1.27$, $p=0.01$), defamation ($M_m=-0.07$, $SD_m=1.24$, $M_f=0.36$, $SD_f=1.09$, $p=0.01$), stress ($M_m=0.38$, $SD_m=1.29$, $M_f=0.98$, $SD_f=1.20$, $p=0.00$), and wrong inference ($M_m=0.43$, $SD_m=1.18$, $M_f=0.81$, $SD_f=1.17$, $p=0.03$). 

For demographics, females were more concerned about inaccuracy ($M_m=0.72$, $SD_m=1.22$, $M_f=1.10$, $SD_f=1.09$, $p=0.02$), privacy ($M_m=0.15$, $SD_m=1.20$, $M_f=0.55$, $SD_f=1.13$, $p=0.02$), consistency ($M_m=0.55$, $SD_m=1.10$, $M_f=1.04$, $SD_f=1.02$, $p=0.00$), stereotyping $M_m=0.87$, $SD_m=1.13$, $M_f=1.26$, $SD_f=0.94$, $p=0.01$), reliability ($M_m=0.47$, $SD_m=1.16$, $M_f=0.87$, $SD_f=1.02$, $p=0.01$), defamation ($M_m=0.20$, $SD_m=1.10$, $M_f=0.62$, $SD_f=1.01$, $p=0.01$), and wrong inference($M_m=0.28$, $SD_m=1.12$, $M_f=0.73$, $SD_f=0.98$, $p=0.00$)

For motivation, females were more concerned about inaccuracy $M_m=0.12$, $SD_m=1.25$, $M_f=0.56$, $SD_f=1.19$, $p=0.02$), 

For Creativity and Problem Solving, females are more concerned about inaccuracy ($M_m=0.28$, $SD_m=1.22$, $M_f=0.65$, $SD_f=1.21$, $p=0.04$), privacy ($M_m=-0.30$, $SD_m=1.26$, $M_f=0.15$, $SD_f=1.24$, $p=0.01$), consistency ($M_m=0.12$, $SD_m=1.29$, $M_f=0.59$, $SD_f=1.15$, $p=0.01$), autonomy ($M_m=0.01$, $SD_m=1.25$, $M_f=0.44$, $SD_f=1.27$, $p=0.02$), stress ($M_m=-0.16$, $SD_m=1.22$, $M_f=0.34$, $SD_f=1.29$, $p=0.01$), wrong inference ($M_m=0.07$, $SD_m=1.28$, $M_f=0.59$, $SD_f=1.21$, $p=0.00$) and control ($M_m=-0.42$, $SD_m=1.12$, $M_f=-0.03$, $SD_f=1.18$, $p=0.02$)

For Physical and Cognitive Impairment, females are more concerned about all the concerns ethics ($M_m=0.48$, $SD_m=1.37$, $M_f=1.07$, $SD_f=1.09$, $p=0.001$),
fairness ($M_m=0.56$, $SD_m=1.23$, $M_f=1.10$, $SD_f=1.08$, $p=0.002$),
inaccuracy ($M_m=0.39$, $SD_m=1.28$, $M_f=1.01$, $SD_f=0.98$, $p=0.000$),
privacy ($M_m=0.24$, $SD_m=1.37$, $M_f=0.88$, $SD_f=1.14$, $p=0.001$),
consistency ($M_m=0.38$, $SD_m=1.24$, $M_f=0.89$, $SD_f=1.07$, $p=0.003$),
autonomy ($M_m=0.43$, $SD_m=1.37$, $M_f=0.95$, $SD_f=1.08$, $p=0.004$),
stereotyping ($M_m=0.53$, $SD_m=1.34$, $M_f=1.03$, $SD_f=1.16$, $p=0.007$),
reliability ($M_m=0.30$, $SD_m=1.24$, $M_f=0.78$, $SD_f=1.18$, $p=0.007$),
defamation ($M_m=0.17$, $SD_m=1.28$, $M_f=0.54$, $SD_f=1.16$, $p=0.041$),
harassment ($M_m=0.28$, $SD_m=1.31$, $M_f=0.73$, $SD_f=1.15$, $p=0.013$),
manipulation ($M_m=0.22$, $SD_m=1.32$, $M_f=0.70$, $SD_f=1.16$, $p=0.009$),
stress ($M_m=0.09$, $SD_m=1.30$, $M_f=0.68$, $SD_f=1.17$, $p=0.001$),
wrong inference ($M_m=0.13$, $SD_m=1.30$, $M_f=0.84$, $SD_f=1.25$, $p=0.000$),
control ($M_m=0.46$, $SD_m=1.38$, $M_f=1.02$, $SD_f=1.14$, $p=0.003$)

\paragraph{Age.} We found differences across age groups for the Motivation data type. Concretely,
compared to younger participants, older participants were more concerned about ethics($M_{18\text{-}24}=-0.37$, $SD_{18\text{-}24}=1.16$, $M_{24+}=0.15$, $SD_{24+}=1.26$, $p=0.00$), fairness ($M_{18\text{-}24}=0.70$, $SD_{18\text{-}24}=1.22$, $M_{24+}=0.96$, $SD_{24+}=1.15$, $p=0.03$), inaccuracy ($M_{18\text{-}24}=0.53$, $SD_{18\text{-}24}=1.26$, $M_{24+}=0.85$, $SD_{24+}=1.10$, $p=0.00$), privacy ($M_{18\text{-}24}=0.49$, $SD_{18\text{-}24}=1.38$, $M_{24+}=0.64$, $SD_{24+}=1.23$, $p=0.03$), and Harassment($M_{18\text{-}24}=-0.54$, $SD_{18\text{-}24}=1.2$, $M_{24+}=-0.18$, $SD_{24+}=1.24$, $p=0.046$),  control ($M_{18\text{-}24}=0.65$, $SD_{18\text{-}24}=1.39$, $M_{24+}=0.85$, $SD_{24+}=1.22$, $p=0.01$). 

Similarly for personality traits prediction, older participants were more concerned about Ethics($M_{18\text{-}24}=0.14$, $SD_{18\text{-}24}=1.24$, $M_{24+}=0.60$, $SD_{24+}=1.09$, $p=0.01$), Harassment($M_{18\text{-}24}=-0.28$, $SD_{18\text{-}24}=1.06$, $M_{24+}=0.24$, $SD_{24+}=1.14$, $p=0.00$) and for Physical and Cognitive Impairment, older participants were more concerned about Consistency ($M_{18\text{-}24}=0.39$, $SD_{18\text{-}24}=1.20$, $M_{24+}=0.85$, $SD_{24+}=1.13$, $p=0.01$), Reliability($M_{18\text{-}24}=0.33$, $SD_{18\text{-}24}=1.27$, $M_{24+}=0.74$, $SD_{24+}=1.19$, $p=0.03$), Defamation($M_{18\text{-}24}=0.15$, $SD_{18\text{-}24}=1.26$, $M_{24+}=0.52$, $SD_{24+}=1.19$, $p=0.04$), Harassment($M_{18\text{-}24}=0.30$, $SD_{18\text{-}24}=1.21$, $M_{24+}=0.68$, $SD_{24+}=1.26$, $p=0.04$), and Manipulation($M_{18\text{-}24}=0.20$, $SD_{18\text{-}24}=1.24$, $M_{24+}=0.68$, $SD_{24+}=1.24$, $p=0.01$)

\paragraph{Education levels.} For motivation prediction, we found a few differences where post-graduate students were more concerned about fairness($M_{\text{UG}}=0.14$, $SD_{\text{UG}}=1.22$, $M_{\text{PG}}=0.55$, $SD_{\text{PG}}=1.24$, $p=0.03$), inaccuracy($M_{\text{UG}}=0.21$, $SD_{\text{UG}}=1.25$, $M_{\text{PG}}=0.58$, $SD_{\text{PG}}=1.22$, $p=0.04$).

\paragraph{Race.} We found interesting differences in harm perceptions between white and non-white participants. White participants, compared to non-white participants, were more concerned for physical or cognitive disability inference due to manipulation ($M_{\text{White}}=0.63$, $SD_{\text{White}}=1.27$, $M_{\text{Non-White}}=0.23$, $SD_{\text{Non-White}}=1.21$, $p=0.03$), stereotyping ($M_{\text{White}}=0.97$, $SD_{\text{White}}=1.24$, $M_{\text{Non-White}}=0.51$, $SD_{\text{Non-White}}=1.27$, $p=0.01$), and fairness ($M_{\text{White}}=1.00$, $SD_{\text{White}}=1.17$, $M_{\text{Non-White}}=0.59$, $SD_{\text{Non-White}}=1.18$, $p=0.02$). 

In contrast, for motivation, not only differing in magnitude of perceived harms, we also found opposite overall trends (positive vs. negative means) between these two groups: non-white participants indicated harms due to reliability($M_{\text{White}}=-0.26$, $SD_{\text{White}}=1.24$, $M_{\text{Non-White}}=0.20$, $SD_{\text{Non-White}}=1.29$, $p=0.02$) and wrong inference ($M_{\text{White}}=-0.04$, $SD_{\text{White}}=1.23$, $M_{\text{Non-White}}=0.40$, $SD_{\text{Non-White}}=1.10$, $p=0.01$), and control ($M_{\text{White}}=-0.70$, $SD_{\text{White}}=1.22$, $M_{\text{Non-White}}=-0.19$, $SD_{\text{Non-White}}=1.23$, $p=0.01$), while white participants did not.


\paragraph{Study disciplines.}
In case of personality prediction, STEM majors were more concerned due to inaccuracy($M_{\text{STEM}}=1.03$, $SD_{\text{STEM}}=0.92$, $M_{\text{Non-STEM}}=0.72$, $SD_{\text{Non-STEM}}=1.11$, $p=0.04$). and in case of Physical or Cognitive Impairment prediction STEM majors were more concerned due to Autonomy ($M_{\text{STEM}}=0.91$, $SD_{\text{STEM}}=1.15$, $M_{\text{Non-STEM}}=0.52$, $SD_{\text{Non-STEM}}=1.32$, $p=0.03$).

These results align with prior research. For example, studies have shown that females and older adults are more concerned about privacy~\cite{prince2023privacy} and discrimination~\cite{pierson2017gender}, as well as overall more negative toward AI decision-making~\cite{make6010017}, when compared to males and younger adults, respectively. Non-white participants' concerns about inferring and using motivation centered around how accurately and reliably it can be inferred, possibly due to them having historically been marked as lazy and intellectually less capable than white people~\cite{diangelo2022whitefragility}. In contrast, white participants perceived greater harms when physical/cognitive disability is inferred, which might seem surprising. One reason could be that, while the prevalence of disability is roughly the same in white and non-white populations~\footnote{\url{https://nces.ed.gov/fastfacts/display.asp?id=60}}, the former group may encounter stereotyping or discrimination at a higher rate for disability status than the latter group (e.g., who faces discrimination for many other reasons).

\subsubsection{Employment context}
Similar to the education context, participants across demographics differed in how they perceived harms in the employment context. Yet, the specific demographic factors across which they differed were not the same, as detailed below. 

\paragraph{Gender.} We did not find any significant differences between male and female participants in the hiring context.

\paragraph{Age.} For emotional state prediction, older participants perceived greater harms due to ethics($M_{18\text{-}24}=0.29$, $SD_{18\text{-}24}=1.25$, $M_{24+}=0.69$, $SD_{24+}=1.17$, $p=0.03$), fairness ($M_{18\text{-}24}=0.48$, $SD_{18\text{-}24}=1.18$, $M_{24+}=1.01$, $SD_{24+}=0.88$, $p=0.00$), inaccuracy ($M_{18\text{-}24}=0.45$, $SD_{18\text{-}24}=1.22$, $M_{24+}=0.96$, $SD_{24+}=0.91$, $p=0.00$), consistency ($M_{18\text{-}24}=0.57$, $SD_{18\text{-}24}=1.13$, $M_{24+}=0.91$, $SD_{24+}=0.91$, $p=0.03$), stereotyping ($M_{18\text{-}24}=0.34$, $SD_{18\text{-}24}=1.18$, $M_{24+}=0.77$, $SD_{24+}=1.03$, $p=0.01$), reliability ($M_{18\text{-}24}=0.44$, $SD_{18\text{-}24}=1.18$, $M_{24+}=0.88$, $SD_{24+}=1.00$, $p=0.01$), defamation ($M_{18\text{-}24}=-0.21$, $SD_{18\text{-}24}=1.09$, $M_{24+}=0.28$, $SD_{24+}=1.16$, $p=0.00$), harassment ($M_{18\text{-}24}=-0.02$, $SD_{18\text{-}24}=1.21$, $M_{24+}=0.41$, $SD_{24+}=1.11$, $p=0.01$), stress ($M_{18\text{-}24}=0.24$, $SD_{18\text{-}24}=1.26$, $M_{24+}=0.62$, $SD_{24+}=1.11$, $p=0.03$), wrong inference ($M_{18\text{-}24}=0.39$, $SD_{18\text{-}24}=1.22$, $M_{24+}=0.84$, $SD_{24+}=0.94$, $p=0.00$), and control ($M_{18\text{-}24}=-0.22$, $SD_{18\text{-}24}=1.23$, $M_{24+}=0.17$, $SD_{24+}=1.22$, $p=0.03$). 

Older adults also perceived greater harms for the inference of personality traits due to ethics($M_{18\text{-}24}=-0.26$, $SD_{18\text{-}24}=1.16$, $M_{24+}=0.09$, $SD_{24+}=1.12$, $p=0.04$), inaccuracy ($M_{18\text{-}24}=0.48$, $SD_{18\text{-}24}=1.13$, $M_{24+}=0.79$, $SD_{24+}=0.95$, $p=0.04$), and reliability ($M_{18\text{-}24}=0.05$, $SD_{18\text{-}24}=1.11$, $M_{24+}=0.40$, $SD_{24+}=1.16$, $p=0.04$) and for the inference of creativity and problem solving due to ethics ($M_{18\text{-}24}=-0.90$, $SD_{18\text{-}24}=1.10$, $M_{24+}=-0.47$, $SD_{24+}=1.19$, $p=0.01$),
fairness ($M_{18\text{-}24}=-0.24$, $SD_{18\text{-}24}=1.29$, $M_{24+}=0.18$, $SD_{24+}=1.23$, $p=0.02$),
stereotyping ($M_{18\text{-}24}=-0.50$, $SD_{18\text{-}24}=1.19$, $M_{24+}=-0.12$, $SD_{24+}=1.16$, $p=0.03$) and for Physical or Cognitive Impairment due to ethics ($M_{18\text{-}24}=0.37$, $SD_{18\text{-}24}=1.41$, $M_{24+}=0.91$, $SD_{24+}=1.10$, $p=0.00$),
fairness ($M_{18\text{-}24}=0.70$, $SD_{18\text{-}24}=1.22$, $M_{24+}=1.06$, $SD_{24+}=1.05$, $p=0.03$),
inaccuracy ($M_{18\text{-}24}=0.50$, $SD_{18\text{-}24}=1.19$, $M_{24+}=0.97$, $SD_{24+}=0.95$, $p=0.00$),
privacy ($M_{18\text{-}24}=0.32$, $SD_{18\text{-}24}=1.40$, $M_{24+}=0.70$, $SD_{24+}=1.20$, $p=0.04$),
consistency ($M_{18\text{-}24}=0.34$, $SD_{18\text{-}24}=1.21$, $M_{24+}=0.89$, $SD_{24+}=1.05$, $p=0.00$),
autonomy ($M_{18\text{-}24}=0.35$, $SD_{18\text{-}24}=1.32$, $M_{24+}=0.72$, $SD_{24+}=1.13$, $p=0.04$),
stereotyping ($M_{18\text{-}24}=0.71$, $SD_{18\text{-}24}=1.25$, $M_{24+}=1.21$, $SD_{24+}=0.94$, $p=0.00$),
reliability ($M_{18\text{-}24}=0.29$, $SD_{18\text{-}24}=1.35$, $M_{24+}=0.66$, $SD_{24+}=1.20$, $p=0.05$),
stress ($M_{18\text{-}24}=0.13$, $SD_{18\text{-}24}=1.27$, $M_{24+}=0.50$, $SD_{24+}=1.20$, $p=0.04$),
wrong inference ($M_{18\text{-}24}=0.12$, $SD_{18\text{-}24}=1.31$, $M_{24+}=0.71$, $SD_{24+}=1.12$, $p=0.00$),
control ($M_{18\text{-}24}=0.60$, $SD_{18\text{-}24}=1.37$, $M_{24+}=1.09$, $SD_{24+}=1.01$, $p=0.00$)

\paragraph{Education levels.} In case of demographics prediction, postgraduates were more concerned due to autonomy ($M_{\text{UG}}=0.32$, $SD_{\text{UG}}=1.11$, $M_{\text{PG}}=0.71$, $SD_{\text{PG}}=1.15$, $p=0.02$), manipulation ($M_{\text{UG}}=0.32$, $SD_{\text{UG}}=1.12$, $M_{\text{PG}}=0.81$, $SD_{\text{PG}}=1.10$, $p=0.00$), and stress ($M_{\text{UG}}=0.11$, $SD_{\text{UG}}=1.13$, $M_{\text{PG}}=0.49$, $SD_{\text{PG}}=1.15$, $p=0.02$).
 
Likewise, post-graduates were more concerned for personality due to fairness($M_{\text{UG}}=0.02$, $SD_{\text{UG}}=1.13$, $M_{\text{PG}}=0.29$, $SD_{\text{PG}}=1.08$, $p=0.10$), privacy ($M_{\text{UG}}=-0.28$, $SD_{\text{UG}}=1.17$, $M_{\text{PG}}=0.24$, $SD_{\text{PG}}=1.13$, $p=0.00$), and manipulation ($M_{\text{UG}}=0.06$, $SD_{\text{UG}}=1.17$, $M_{\text{PG}}=0.39$, $SD_{\text{PG}}=1.03$, $p=0.05$). In addition, postgraduates were more concerned for motivation due to ethics ($M_{\text{UG}}=-0.61$, $SD_{\text{UG}}=1.09$, $M_{\text{PG}}=-0.08$, $SD_{\text{PG}}=1.07$, $p=0.00$), autonomy ($M_{\text{UG}}=-0.39$, $SD_{\text{UG}}=1.11$, $M_{\text{PG}}=-0.01$, $SD_{\text{PG}}=1.22$, $p=0.03$), privacy($M_{\text{UG}}=-0.72$, $SD_{\text{UG}}=0.95$, $M_{\text{PG}}=-0.31$, $SD_{\text{PG}}=1.27$, $p=0.09$), and reliability ($M_{\text{UG}}=-0.25$, $SD_{\text{UG}}=1.14$, $M_{\text{PG}}=0.12$, $SD_{\text{PG}}=1.28$, $p=0.04$). For emotional state prediction due to autonomy ($M_{UG}=0.18$, $SD_{UG}=1.20$, $M_{PG}=0.53$, $SD_{PG}=1.15$, $p=0.05$). For creativity and problem solving due to ethics ($M_{UG}=-0.81$, $SD_{UG}=1.08$, $M_{Grad+}=-0.42$, $SD_{Grad+}=1.26$, $p=0.02$), and
control ($M_{UG}=-0.78$, $SD_{UG}=1.08$, $M_{Grad+}=-0.37$, $SD_{Grad+}=1.27$, $p=0.02$). For physical and cognitive impairment due to inaccuracy ($M_{UG}=0.62$, $SD_{UG}=1.14$, $M_{Grad+}=1.00$, $SD_{Grad+}=0.99$, $p=0.02$),
consistency ($M_{UG}=0.49$, $SD_{UG}=1.21$, $M_{Grad+}=0.90$, $SD_{Grad+}=1.03$, $p=0.01$),
manipulation ($M_{UG}=0.35$, $SD_{UG}=1.19$, $M_{Grad+}=0.71$, $SD_{Grad+}=1.05$, $p=0.03$),
wrong inference ($M_{UG}=0.30$, $SD_{UG}=1.27$, $M_{Grad+}=0.66$, $SD_{Grad+}=1.18$, $p=0.05$),
control ($M_{UG}=0.70$, $SD_{UG}=1.27$, $M_{Grad+}=1.10$, $SD_{Grad+}=1.05$, $p=0.02$)

\paragraph{Race.} Similar to the education context, we found most diverse harm perceptions between white and non-white participants in employment context. White participants perceived greater harms associated with emotion inference due to fairness ($M_{\text{White}}=0.93$, $SD_{\text{White}}=0.99$, $M_{\text{Non-White}}=0.59$, $SD_{\text{Non-White}}=1.12$, $p=0.03$) and inaccuracy ($M_{\text{White}}=0.87$, $SD_{\text{White}}=0.99$, $M_{\text{Non-White}}=0.55$, $SD_{\text{Non-White}}=1.17$, $p=0.04$). More, white participants were more concerned for physical or cognitive impairment due to privacy ($M_{White}=0.75$, $SD_{White}=1.30$, $M_{NonWhite}=0.30$, $SD_{NonWhite}=1.27$, $p=0.02$),
consistency ($M_{White}=0.81$, $SD_{White}=1.07$, $M_{NonWhite}=0.48$, $SD_{NonWhite}=1.21$, $p=0.04$),
defamation ($M_{White}=0.56$, $SD_{White}=1.10$, $M_{NonWhite}=0.15$, $SD_{NonWhite}=1.19$, $p=0.01$),
harassment ($M_{White}=0.79$, $SD_{White}=1.18$, $M_{NonWhite}=0.41$, $SD_{NonWhite}=1.28$, $p=0.04$),
wrong inference ($M_{White}=0.71$, $SD_{White}=1.17$, $M_{NonWhite}=0.17$, $SD_{NonWhite}=1.25$, $p=0.00$)

We again found opposite trends (positive vs. negative means) between these two groups. For motivation, white participants had opposite risk perception regarding wrong inference ($M_{\text{White}}=0.37$, $SD_{\text{White}}=1.18$, $M_{\text{Non-White}}=-0.30$, $SD_{\text{Non-White}}=2.61$, $p=0.01$) and reliability ($M_{\text{White}}=0.08$, $SD_{\text{White}}=1.21$, $M_{\text{Non-White}}=-0.33$, $SD_{\text{Non-White}}=1.20$, $p=0.02$). Similar result was observed for harms from wrong inference ($M_{\text{White}}=0.07$, $SD_{\text{White}}=1.17$, $M_{\text{Non-White}}=-0.29$, $SD_{\text{Non-White}}=1.21$, $p=0.04$) when inferring creativity. 

Finally, results were mixed for 
personality traits prediction. While white participants feared greater risks due to harassment($M_{\text{White}}=-0.25$, $SD_{\text{White}}=1.13$, $M_{\text{Non-White}}=0.12$, $SD_{\text{Non-White}}=1.19$, $p=0.03$) than non-white participants, the latter group feared greater risks due to wrong inference ($M_{\text{White}}=0.45$, $SD_{\text{White}}=1.07$, $M_{\text{Non-White}}=0.09$, $SD_{\text{Non-White}}=1.10$, $p=0.02$) for the same data type.

\paragraph{Study discipline.}
In case of creativity and problem solving prediction Non-STEM major participants were more concerned due to reliability($M_{\text{STEM}}=-0.51$, $SD_{\text{STEM}}=1.23$, $M_{\text{Non-STEM}}=-0.13$, $SD_{\text{Non-STEM}}=1.23$, $p=0.03$).

Overall, again we see numerous variations in how different sub-populations perceive privacy harms, which we hypothesize has been affected by social structure and their lived experiences. Worth noting here is that, in the employment context, we see differences between undergraduate and postgraduate participants, possibly due to the latter group having more frequent first-hand experience with AI in employment (while there was no such differences in the education context).

\begin{SummaryBox}{Takeaways}
    \begin{enumerate}
        \item Privacy harm perceptions across demographic factors are highly nuanced and diverse, possibly reflecting historical socio-political factors and lived experiences.
        \item Surfacing diversity in perceived harms from a given technology can guide mitigating techniques, e.g., an algorithm deployed at a university primarily serving a specific demographic may prioritize mitigating harms to that population.
    \end{enumerate}
\end{SummaryBox}

\subsection{External validity}
We demonstrate the external validity of the harms we study by applying them to label real-world incidents involving AI. We use the AIID (AI Incident Database) that documents hundreds of incidents involving the use, misuse, or failure of AI systems across a variety of sectors (e.g., education, employment, law enforcement, and healthcare). First, the lead author identified
incidents in education and employment contexts, where the affected users, decision targets, or environments involved students, teachers, job seekers, employees, or related institutional settings. We found 36 such incidents (out of more than 200 incidents). Then, the lead author reviewed each incident description in the database as well as additional materials (notes, media coverage, and reported outcomes) and tagged it with applicable harm categories. Of these, 35 incidents were successfully mapped to at least one of the 14 concerns. One incident was excluded due to insufficient information for reliable coding. Another author reviewed the annotated data; minor disagreements and unclarity were resolved through discussions, but there was no major change in the labeling.

Table~\ref{tab:harms_by_context} shows the number of incidents labeled with each harm type, along with brief descriptions of the incidents. Note that one incident could have multiple tags (e.g., a school surveillance system could raise both privacy and autonomy concerns). The successful application of the harm categories to label all (except one) of the incidents the validity of our harm taxonomy and its applicability to categorize diverse real-world incidents, beyond the two decision contexts we studied.

\begin{table*}[h!]
\centering
\footnotesize
\caption{AI-Related Incidents by Harm Type in Education and Employment Contexts from Harms Database}
\resizebox{\textwidth}{!}{
\begin{tabular}{l r p{5.2cm} p{8 cm}}
\hline
\textbf{Harm Type} & \textbf{Count} & \textbf{Incident Names} & \textbf{Short Description and Context Reasoning} \\
\hline
Privacy & 12 & \textit{Anderstorp HS, Nice/Marseille Schools, Lockport CSD, Intel Emotion AI, SG Mindline, Victoria LoopLearn, Nanjing Univ., Niulanshan, Hangzhou HS, Turnitin AI, Poland PSZ, Lancaster HS, Westfield HS} & AI systems captured or misused biometric/audio/textual data in schools or employment agencies. Education examples include facial ID, attention tracking; employment cases include profiling tools. \\

Ethics & 9 & \textit{Anderstorp HS, Nice/Marseille Schools, Lockport CSD, Victoria LoopLearn, Carmel HS, Baltimore HS, Lancaster HS, Westfield HS, Austria AMS} & These incidents involved ethically questionable use of AI such as surveillance of minors, use of deepfakes, or unfair labor tools. Sector classification is based on setting (schools, job services). \\

Fairness & 8 & \textit{Lockport CSD, Hangzhou HS, DC Teacher Eval, Turnitin AI, Austria AMS, Workday AI, Poland PSZ, Michigan MiDAS} & Algorithms exhibited racial, gender, or economic bias in education (student systems, teacher evaluation) and employment (hiring, unemployment fraud systems). \\

Inaccuracy & 8 & \textit{Lockport CSD, Intel Emotion AI, SG Mindline, Niulanshan, Hangzhou HS, DC Teacher Eval, Turnitin AI, Michigan MiDAS} & AI outputs were unreliable or incorrect, affecting grades, employment eligibility, or system performance. Identified from sector and use-case context. \\

Stress & 4 & \textit{SG Mindline, Lancaster HS, Westfield HS, Michigan MiDAS} & Psychological distress arose due to AI-generated harassment (education) or wrongful fraud accusations (employment). \\

Harassment & 4 & \textit{Carmel HS, Baltimore HS, Lancaster HS, Westfield HS} & Victims were harassed via deepfakes and false audio in educational institutions. \\

Stereotyping & 3 & \textit{Carmel HS, Austria AMS, Workday AI} & Use of AI reflected or perpetuated age, race, or ability-based stereotypes. Mapped to context via actor involved (student vs. applicant). \\

Wrong Inference & 3 & \textit{Carmel HS, Baltimore HS, Austria AMS} & AI systems made unjustified claims (e.g., speech attribution, work potential), leading to harm. Context inferred from case setting. \\

Autonomy & 2 & \textit{Anderstorp HS, Nice/Marseille Schools} & Students and families had no choice regarding AI surveillance in school pilots. \\

Reliability & 2 & \textit{Intel Emotion AI, SG Mindline} & Systems failed to consistently deliver promised performance, especially in educational use. \\

Defamation & 2 & \textit{Carmel HS, Baltimore HS} & Deepfakes falsely attributed speech/behavior, damaging reputations in school settings. \\

Control & 1 & \textit{Anderstorp HS} & Parents and students had no opt-out or control over AI deployments in schools. \\

Manipulation & 1 & \textit{Carmel HS} & AI was intentionally used by students to deceive others (e.g., fake racist video). \\

Consistency & 0 & -- &  -- \\
\hline
\end{tabular}
}
\label{tab:harms_by_context}
\end{table*}

%% file: sections/cronbach_alpha_table.tex
\begin{table*}[h!]
\centering
\caption{Cronbach's Alpha (leave-one-out) and Item-Total Correlation (in parentheses) by Attribute for Education Context}
\begin{tabularx}{\textwidth}{l *{6}{>{\centering\arraybackslash}X}}
\toprule
\textbf{Item Removed} & \textbf{Demographics} & \textbf{Personality} & \textbf{Motivation} & \textbf{Emotion} & \textbf{Creativity} & \textbf{Impairment} \\
\midrule
- & 0.93 & 0.94 & 0.95 & 0.95 & 0.96 & 0.96 \\
unethical & 0.92 (0.76) & 0.93 (0.77) & 0.94 (0.83) & 0.94 (0.76) & 0.95 (0.84) & 0.96 (0.81) \\
biased & 0.92 (0.75) & 0.93 (0.74) & 0.94 (0.77) & 0.94 (0.79) & 0.95 (0.75) & 0.96 (0.83) \\
inaccurate & 0.92 (0.74) & 0.93 (0.74) & 0.94 (0.76) & 0.94 (0.78) & 0.95 (0.78) & 0.96 (0.80) \\
privacy & 0.93 (0.57) & 0.93 (0.72) & 0.94 (0.78) & 0.95 (0.70) & 0.95 (0.77) & 0.96 (0.74) \\
inconsistency & 0.92 (0.72) & 0.93 (0.71) & 0.94 (0.75) & 0.94 (0.79) & 0.95 (0.80) & 0.96 (0.81) \\
freedom & 0.92 (0.68) & 0.93 (0.76) & 0.94 (0.73) & 0.94 (0.76) & 0.95 (0.80) & 0.96 (0.78) \\
stereotyping & 0.92 (0.72) & 0.93 (0.71) & 0.95 (0.70) & 0.95 (0.67) & 0.95 (0.77) & 0.96 (0.80) \\
reliable & 0.92 (0.64) & 0.93 (0.64) & 0.94 (0.73) & 0.94 (0.76) & 0.95 (0.79) & 0.96 (0.76) \\
defamation & 0.92 (0.71) & 0.93 (0.73) & 0.94 (0.74) & 0.94 (0.73) & 0.96 (0.71) & 0.96 (0.75) \\
harassment & 0.92 (0.63) & 0.94 (0.59) & 0.95 (0.71) & 0.95 (0.67) & 0.95 (0.77) & 0.96 (0.78) \\
manipulation & 0.92 (0.61) & 0.93 (0.71) & 0.94 (0.73) & 0.94 (0.79) & 0.95 (0.75) & 0.96 (0.78) \\
stress & 0.93 (0.53) & 0.93 (0.66) & 0.95 (0.68) & 0.95 (0.66) & 0.96 (0.73) & 0.96 (0.74) \\
inference & 0.92 (0.63) & 0.93 (0.72) & 0.95 (0.72) & 0.94 (0.81) & 0.95 (0.79) & 0.96 (0.77) \\
control & 0.92 (0.64) & 0.94 (0.55) & 0.95 (0.62) & 0.95 (0.59) & 0.96 (0.70) & 0.96 (0.74) \\
\bottomrule
\end{tabularx}
\label{tab:education_alpha_itc}
\end{table*}

%% file: sections/5_discussions.tex
\section{Discussions and conclusions}
In this paper, we proposed and operationalized a harm-centric conceptualization of privacy by investigating perceived privacy harms in the context of using AI models in education and employment. Below, we discuss the theoretical and practical implications of this research.

\paragraph{Advancing the understanding of privacy.} A general definition of privacy that captures all its aspects remains elusive~\cite{soloveTaxonomyPrivacy}. Several conceptualizations of privacy have emerged over the decades; each of which has increasingly clarified our understanding of privacy and contributed to technical innovations in protecting it. Here, we propose a harm-centric view that leverages the conceptualization of privacy as a shield from different harms that can arise due to personal data use. We empirically demonstrate the reliability and validity of measuring privacy this way, which proves that this conceptualization can be used in other contexts to empirically understand privacy.

\paragraph{Nuanced discovery of privacy harms through the lens of procedural justice.} Our approach surfaced privacy harms that may go undetected by other frameworks. Our data shows that people anticipate privacy harms even if the data is not private or there is no direct data leak or malicious adversary. Importantly, using the lens of procedural justice enabled us to surface privacy harms from otherwise justifiable or even desirable systems. For example, one could argue that automation would expedite identifying students at risk of dropping out and providing targeted assistance, which is desirable; however, procedural justice requires that the operating procedure of that system does not harm the affected population. Indeed, our data shows that participants feared being discriminated against by the algorithms because of how they work (e.g., implicitly relying on protected attributes), even if the outcome is useful. 

\paragraph{Equitable privacy with informed prioritization.} Our data show significant variances in how sub-groups of a population vary in their perception of privacy harms. Not only did participants vary along the degree of harm, but different sub-groups also anticipated being impacted by different types of harms in a given context. This allows us to identify the group likely to be most impacted, even if it's a minority. This also may provide clarity on how to prioritize among mitigating harms when needed; for example, when simultaneously satisfying different criteria like accuracy, privacy, and fairness may be computationally infeasible~\cite{guPrivacyAccuracyModel-2022, changPrivacyRisksAlgorithmic-2021}.

\paragraph{Threat modeling.} Understanding how personal data use harms data subjects can be useful in threat modeling for a system that uses such data. For example, Massey~\etal used Solove's taxonomy~\cite{soloveTaxonomyPrivacy} for mapping system vulnerabilities to privacy-violating actions~\cite{massey_requirements-based_2008}. Recent threat modeling frameworks have started to go beyond threats from malicious users or active adversaries, and include harms to data subjects, both intentional and unintentional~\cite{landuytPrivacyImpactTree-2025}. There can be a direct mapping from privacy harms to the threats a system poses to its users; thus, studying privacy following the harm-centered approach can greatly accelerate user-centered threat modeling.

\paragraph{Recommendations to improve privacy.} Empirical insights from our study could guide concrete policy and technical interventions to improve privacy. For example, we find that participants expressed greater levels of concern when certain data (e.g., demographics) is used in the education context than in the employment context. As our literature review shows, these data are widely used both in research and commercial learning analytics tools to predict many outcomes about students. Fortunately, institutional policies largely dictate how students' data can be used, both internally and by external entities such as technology providers~\cite{kelso_trust_2024}. Thus, institutes could implement preventive measures, such as prohibiting the repurposing of trained models (to infer personal data) by, e.g., extracting internal representation on input data. Additionally, both educational institutions and commercial entities can use technical mechanisms such as adversarial censoring to remove problematic personal data embedded in trained models~\cite{edtech-pets22}.

\paragraph{Future usable privacy research.} A long-term goal of usable privacy research has been to link people's privacy mental model and how that influences their privacy behaviors (or at least intention to behave in a privacy-protective manner). This has proven to be difficult, and in particular, we still lack an understanding of how people conceptualize privacy and related constructs (e.g., privacy concerns) and how to reliably measure them~\cite{colnago-concern-preference}. Contextual investigation of harm-centric conceptualization of privacy seems promising to shed light on when people may desire privacy protection. Future works could study how well that desire for privacy leads to behaviors that are protective of privacy.   

To conclude, we demonstrated that measuring perceived harms from personal data use can be a reliable and valid quantification of people's conceptualization privacy as a shield against such harms. We hope that future research will adopt this formulation and approach to understand privacy in other domains.

%% file: sections/appendix.tex
\appendix
\section{Appendix}
\subsection{Survey questionnaire}
\label{sec:ques}
\subsubsection{Instructions}
\noindent \textbf{Please read the following paragraph carefully before answering any questions. }

(\textbf{Education context}) Many schools now use artificial intelligence-based (AI) tools to support student success, which might expedite the process and enable evaluating student data at scale. AI tools can collect and analyze a lot of data from different sources in a much shorter time compared to human educators. For example, an AI tool may analyze your academic records, interactions with learning management systems, campus mobility data, as well as profiles and activities on external sites (e.g., GitHub or personal websites). Using all these data it can infer different criteria and based on which it decides whether you might be at risk of dropping out of a course.

Considering this situation, please answer the following questions about your opinion on AI tools predicting information about you.

(\textbf{Hiring context}) Artificial intelligence-based (AI) tools are being used for recruitment, which might enable evaluating job applications faster. AI tools can collect and analyze a lot of data from different sources in a much shorter time compared to human recruiters. For example, an AI tool may analyze your resume, information available online (e.g., from personal websites or Github profiles), recorded interviews, as well as data collected by different tools you used for studying (including learning management systems and other apps) to infer certain attributes about you. It can then use all these data to quantify different criteria and based on them it decides whether you should be employed. 

Considering this situation, please answer the following questions about your opinion on AI recruitment tools predicting information about you.

\subsubsection{Questions}
After the general (context-specific) instruction, we explained the three attributes (shown below), after each of the explanations, we asked the participants to rate the following statements using a 5-point Likert scale (``Strongly agree'' to ``Strongly disagree''). Each participant only answered questions about one decision context: dropout risk or job fit assessment. The [Attribute] field was populated with the three attributes listed below.

\begin{enumerate}
    \item Use of [Attribute] to decide [dropout risk | job fit] is unethical. 
    \item Use of [Attribute] to decide [dropout risk | job fit] can lead to biased decisions.
    \item Use of [Attribute] to decide [dropout risk | job fit] can lead to inaccurate decisions.
    \item Use of [Attribute] to decide [dropout risk | job fit] violates my privacy. 
    \item Use of [Attribute] to decide [dropout risk | job fit] can lead to inconsistency in the decision-making process.
    \item Use of [Attribute] to decide [dropout risk | job fit] goes against my freedom to choose how I want to be judged. 
    \item Use of [Attribute] to decide [dropout risk | job fit] can lead to stereotyping. 
    \item {[Attribute]} is not a reliable predictor of [dropout risk | job fit]. 
    \item Use of [Attribute] to decide [dropout risk | job fit] can lead to defamation. 
    \item Use of [Attribute] to decide [dropout risk | job fit] can lead to harassment. 
    \item Use of [Attribute] to decide [dropout risk | job fit] can lead to manipulation. 
    \item Use of [Attribute] to decide [dropout risk | job fit] will create stress for me. 
    \item {[Attribute]} cannot be correctly inferred to use it to decide [dropout risk | job fit]. 
    \item Use of [Attribute] is inappropriate since I have no control over [Attribute].
\end{enumerate}

\subsubsection{Attribute descriptions}
\hspace*{\parindent}

\begin{itemize}
        \item \textbf{Demographics:} Demographic information, such as age, gender, and race, can be predicted using machine learning algorithms based on various sources of information, including writing style, images/videos, and how users interact with online platforms.

    \item \textbf{Personality Traits:} Personality traits, such as extroversion, conscientiousness, and openness, can be predicted using machine learning algorithms based on various sources of information, including social media activity, language use in communications, Video Interviews and Github Issues and Commits Activity.

    \item \textbf{Emotional State:} Emotional state, such as stress levels and emotional stability, can be predicted using machine learning algorithms based on various sources of information, including facial expressions in videos, tone of voice in audio recordings, and language use in written communications.
    
    \end{itemize}

\subsubsection{Demographic questions}
Finally, we asked questions about gender, age, current education level, and race.

\subsection{Additional results}
\begin{figure*}[h!]
    \centering
    \includegraphics[width=0.98\textwidth, height=0.97\textheight, keepaspectratio]{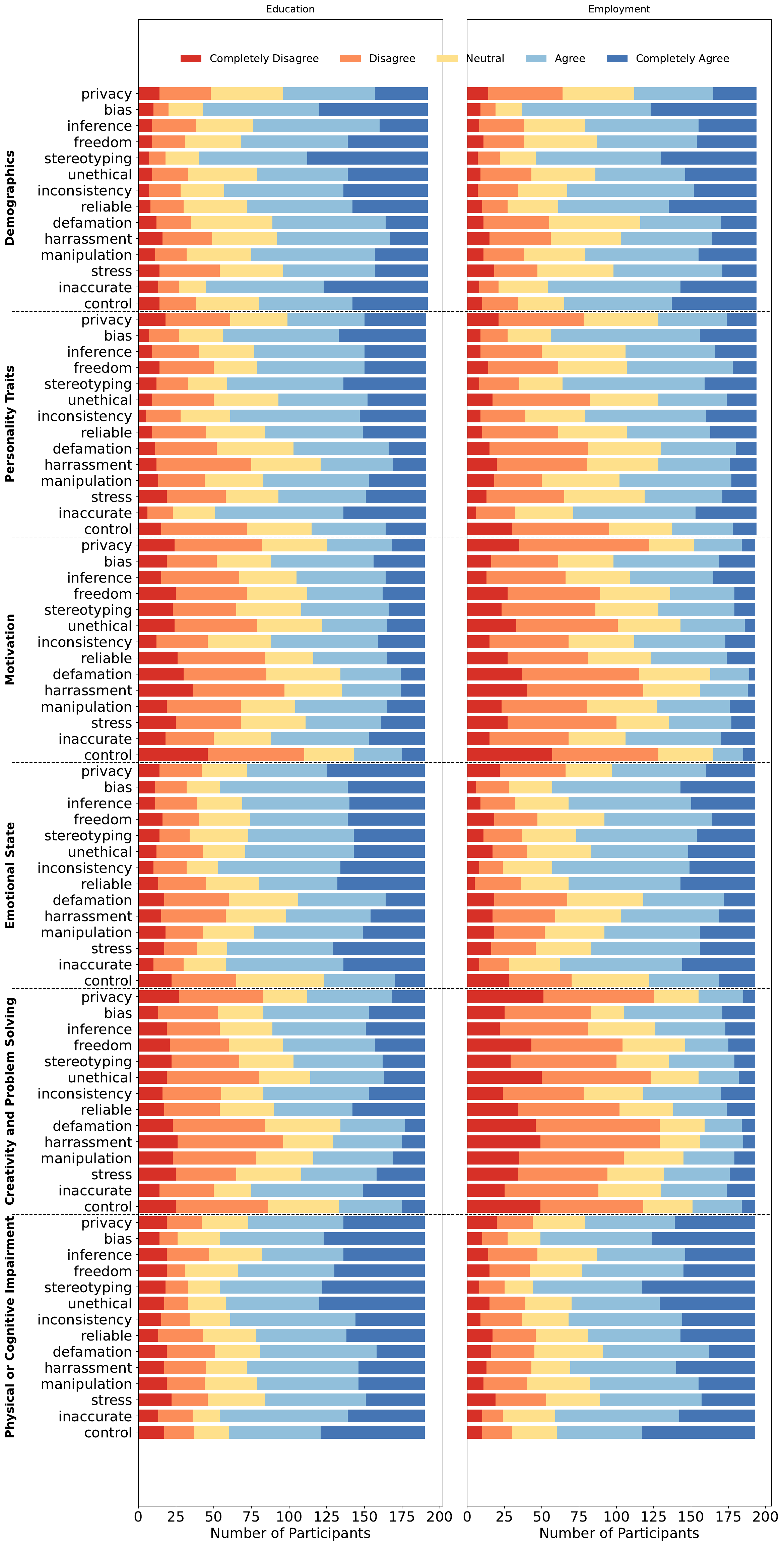}
    \caption{Distribution of participants' responses to harm statements (left: education, right: employment).} 
    \label{fig:response_counts}
\end{figure*}

\input{sections/harms_database}
\input{sections/Categorical_comparison}

%% file: sections/Categorical_comparison.tex
\begin{table}[htbp]
\centering
\caption{Number of participants when grouped by demographic factors}
\begin{tabular}{l r r}
\toprule
\textbf{Grouping} & \textbf{Education context} & \textbf{Employment context} \\
 \midrule
White & 112 & 107 \\

Non-White & 75 & 82 \\
 
STEM & 90 & 99 \\
 
Non-STEM & 110 & 101 \\
 
Male & 89 & 84 \\
 
Female& 97 & 102 \\
 
Undergrad & 96 & 105 \\
 
Post-grad & 86 & 83 \\
 
18--24 years old & 79 & 82 
\\
 
$>$24 years old & 110 & 107 \\
\bottomrule
\end{tabular}
\label{tab:categroical_comparison}
\end{table}

%% file: template.bbl
\begin{thebibliography}{10}

\bibitem{xuExaminingFormationIndividuals-2008}
Heng Xu, Tamara Dinev, H.~Smith, and Paul Hart.
\newblock Examining the {Formation} of {Individual}'s {Privacy} {Concerns}: {Toward} an {Integrative} {View}.
\newblock {\em ICIS 2008 Proceedings}, January 2008.

\bibitem{soloveDataWhatDataDoes2024}
Daniel~J. Solove.
\newblock Data {Is} {What} {Data} {Does}: {Regulating} {Based} on {Harm} and {Risk} {Instead} of {Sensitive} {Data}, January 2024.

\bibitem{dinevExtendedPrivacyCalculusECommerce2006}
Tamara Dinev and Paul Hart.
\newblock An {Extended} {Privacy} {Calculus} {Model} for {E}-{Commerce} {Transactions}.
\newblock {\em Information Systems Research}, 17(1):61--80, March 2006.
\newblock Publisher: INFORMS.

\bibitem{soloveMythofPrivacyParadox2021}
Daniel~J. Solove.
\newblock The {Myth} of the {Privacy} {Paradox}.
\newblock {\em George Washington Law Review}, 89:1, 2021.

\bibitem{solove_murky_2023}
Daniel~J. Solove.
\newblock Murky {Consent}: {An} {Approach} to the {Fictions} of {Consent} in {Privacy} {Law}.
\newblock {\em SSRN Electronic Journal}, 2023.

\bibitem{nissenbaum2004CI}
Helen Nissenbaum.
\newblock Privacy as {Contextual} {Integrity}.
\newblock 79:119, 2004.
\newblock Publisher: HeinOnline.

\bibitem{kumarRoadmapApplyingCI2024}
Priya~C. Kumar, Michael Zimmer, and Jessica Vitak.
\newblock A {Roadmap} for {Applying} the {Contextual} {Integrity} {Framework} in {Qualitative} {Privacy} {Research}.
\newblock {\em Proc. ACM Hum.-Comput. Interact.}, 8(CSCW1):219:1--219:29, April 2024.

\bibitem{angel2024distinguishingPrivacy}
Marı́a~P Angel and Ryan Calo.
\newblock Distinguishing privacy law.
\newblock {\em Columbia Law Review}, 124(2):507--562, 2024.
\newblock Publisher: JSTOR.

\bibitem{richardsWhyPrivacyMatters-2021}
Neil~M. Richards.
\newblock Why {Privacy} {Matters}: {An} {Introduction}, December 2021.

\bibitem{calo2014privacyHarmExceptionalizm}
Ryan Calo.
\newblock Privacy harm exceptionalism.
\newblock 12:361, 2014.
\newblock Publisher: HeinOnline.

\bibitem{bhatiaEmpiricalMeasurementPerceived-2018}
Jaspreet Bhatia and Travis~D. Breaux.
\newblock Empirical {Measurement} of {Perceived} {Privacy} {Risk}.
\newblock {\em ACM Trans. Comput.-Hum. Interact.}, 25(6):34:1--34:47, December 2018.

\bibitem{hunkenschroer_is_2023}
Anna~Lena Hunkenschroer and Alexander Kriebitz.
\newblock Is {AI} recruiting (un)ethical? {A} human rights perspective on the use of {AI} for hiring.
\newblock {\em AI and Ethics}, 3(1):199--213, February 2023.

\bibitem{dropouts-moocs}
Wenzheng Feng, Jie Tang, and Tracy~Xiao Liu.
\newblock Understanding dropouts in {MOOCs}.
\newblock {\em Proceedings of the AAAI Conference on Artificial Intelligence}, 33(01):517--524, July 2019.

\bibitem{oulad_dropout_performance_prediction}
Nikhil~Indrashekhar Jha, Ioana Ghergulescu, and Arghir-Nicolae Moldovan.
\newblock {OULAD} {MOOC} dropout and result prediction using ensemble, deep learning and regression techniques.
\newblock In {\em {CSEDU} (2)}, pages 154--164, 2019.

\bibitem{predictive-LA-survey}
Ying Cui, Fu~Chen, Ali Shiri, and Yaqin Fan.
\newblock Predictive analytic models of student success in higher education: {A} review of methodology.
\newblock {\em Information and Learning Sciences}, 2019.
\newblock Publisher: Emerald Publishing Limited.

\bibitem{noauthor_pdf_2024}
({PDF}) {The} {Impact} of {AI} on {Recruitment} and {Selection} {Processes}: {Analysing} the role of {AI} in automating and enhancing recruitment and selection procedures.
\newblock {\em ResearchGate}, October 2024.

\bibitem{noauthor_predictive_nodate}
Predictive {Analytics} for {Employee} {Retention}: {Forecasting} and {Preventing} {Turnover} - {Hirebee}.

\bibitem{kodiyan2019-overview}
Akhil~Alfons Kodiyan.
\newblock An overview of ethical issues in using {AI} systems in hiring with a case study of {Amazon}’s {AI} based hiring tool.
\newblock {\em Researchgate Preprint}, pages 1--19, 2019.

\bibitem{youngInclusiveTechPolicy-2019}
Meg Young, Lassana Magassa, and Batya Friedman.
\newblock Toward inclusive tech policy design: a method for underrepresented voices to strengthen tech policy documents.
\newblock {\em Ethics and Information Technology}, 21(2):89--103, June 2019.

\bibitem{knijnenburg_modern_2022}
Bart~P. Knijnenburg, Xinru Page, Pamela Wisniewski, Heather~Richter Lipford, Nicholas Proferes, and Jennifer Romano, editors.
\newblock {\em Modern {Socio}-{Technical} {Perspectives} on {Privacy}}.
\newblock Springer International Publishing, Cham, 2022.

\bibitem{knowlesUnParadoxingPrivacyConsidering-2023}
Bran Knowles and Stacey Conchie.
\newblock Un-{Paradoxing} {Privacy}: {Considering} {Hopeful} {Trust}.
\newblock {\em ACM Trans. Comput.-Hum. Interact.}, 30(6):87:1--87:24, September 2023.

\bibitem{guberek_keeping_2018}
Tamy Guberek, Allison McDonald, Sylvia Simioni, Abraham~H. Mhaidli, Kentaro Toyama, and Florian Schaub.
\newblock Keeping a {Low} {Profile}? {Technology}, {Risk} and {Privacy} among {Undocumented} {Immigrants}.
\newblock In {\em Proceedings of the 2018 {CHI} {Conference} on {Human} {Factors} in {Computing} {Systems}}, {CHI} '18, pages 1--15, New York, NY, USA, April 2018. Association for Computing Machinery.

\bibitem{palenUnpackingPrivacyNetworked-2003}
Leysia Palen and Paul Dourish.
\newblock Unpacking "privacy" for a networked world.
\newblock In {\em Proceedings of the {SIGCHI} {Conference} on {Human} {Factors} in {Computing} {Systems}}, {CHI} '03, pages 129--136, New York, NY, USA, April 2003. Association for Computing Machinery.

\bibitem{richardsTakingTrustSeriously-2015}
Neil Richards and Woodrow Hartzog.
\newblock Taking {Trust} {Seriously} in {Privacy} {Law}.
\newblock {\em Stanford Technology Law Review}, 19:431, 2015.

\bibitem{hullSuccessfulFailureWhat-2015}
Gordon Hull.
\newblock Successful failure: what {Foucault} can teach us about privacy self-management in a world of {Facebook} and big data.
\newblock {\em Ethics and Information Technology}, 17(2):89--101, June 2015.

\bibitem{soloveKafkaAgeAIPrivacy2024}
Daniel~J. Solove and Woodrow Hartzog.
\newblock Kafka in the {Age} of {AI} and the {Futility} of {Privacy} as {Control}, January 2024.

\bibitem{nissenbaum2009privacy}
Helen Nissenbaum.
\newblock {\em Privacy in context: {Technology}, policy, and the integrity of social life}.
\newblock Stanford University Press, 2009.

\bibitem{nissenbaumContextualIntegrityDataChain2019}
Helen Nissenbaum.
\newblock Contextual {Integrity} {Up} and {Down} the {Data} {Food} {Chain}.
\newblock {\em Theoretical Inquiries in Law}, 20(1):221--256, January 2019.
\newblock Publisher: De Gruyter.

\bibitem{umbachDeepFakePornography2024}
Rebecca Umbach, Nicola Henry, Gemma~Faye Beard, and Colleen~M. Berryessa.
\newblock Non-{Consensual} {Synthetic} {Intimate} {Imagery}: {Prevalence}, {Attitudes}, and {Knowledge} in 10 {Countries}.
\newblock In {\em Proceedings of the {CHI} {Conference} on {Human} {Factors} in {Computing} {Systems}}, {CHI} '24, pages 1--20, New York, NY, USA, May 2024. Association for Computing Machinery.

\bibitem{vandepoelSociallyDisruptiveTechnologies-2022}
Ibo van~de Poel.
\newblock Socially {Disruptive} {Technologies}, {Contextual} {Integrity}, and {Conservatism} {About} {Moral} {Change}.
\newblock {\em Philosophy \& Technology}, 35(3):82, August 2022.

\bibitem{solove2015meaningValue}
Daniel~J Solove.
\newblock The meaning and value of privacy.
\newblock {\em Social dimensions of privacy: Interdisciplinary perspectives}, 71, 2015.
\newblock Publisher: Cambridge University Press Cambridge.

\bibitem{investigate-risk-perception}
Nina Gerber, Benjamin Reinheimer, and Melanie Volkamer.
\newblock Investigating people's privacy risk perception.
\newblock {\em Proc. Priv. Enhancing Technol.}, 2019(3):267--288, 2019.

\bibitem{jakobiItWhatTheyDoWithData2019}
Timo Jakobi, Sameer Patil, Dave Randall, Gunnar Stevens, and Volker Wulf.
\newblock It {Is} {About} {What} {They} {Could} {Do} with the {Data}: {A} {User} {Perspective} on {Privacy} in {Smart} {Metering}.
\newblock {\em ACM Trans. Comput.-Hum. Interact.}, 26(1):2:1--2:44, January 2019.

\bibitem{colnagoThereReversePrivacy-2023}
Jessica Colnago, Lorrie~Faith Cranor, and Alessandro Acquisti.
\newblock Is {There} a {Reverse} {Privacy} {Paradox}? {An} {Exploratory} {Analysis} of {Gaps} {Between} {Privacy} {Perspectives} and {Privacy}-{Seeking} {Behaviors}, July 2023.

\bibitem{karwatzkiYESFIRMSHAVE-2018}
Sabrina Karwatzki, Manuel Trenz, and Daniel Veit.
\newblock {YES}, {FIRMS} {HAVE} {MY} {DATA} {BUT} {WHAT} {DOES} {IT} {MATTER}? {MEASURING} {PRIVACY} {RISKS}.
\newblock {\em Research Papers}, November 2018.

\bibitem{jakobiTaxonomyUserperceivedPrivacy-2022a}
Timo Jakobi, Maximilian von Grafenstein, Patrick Smieskol, and Gunnar Stevens.
\newblock A {Taxonomy} of user-perceived privacy risks to foster accountability of data-based services.
\newblock {\em Journal of Responsible Technology}, 10:100029, July 2022.

\bibitem{soloveTaxonomyPrivacy}
Daniel~J. Solove.
\newblock A {Taxonomy} of {Privacy}.
\newblock {\em University of Pennsylvania Law Review}, 154(3):477--564, 2005.

\bibitem{brooksIntroductionPrivacyEngineering-2017}
Sean Brooks, Michael Garcia, Naomi Lefkovitz, Suzanne Lightman, and Ellen Nadeau.
\newblock An {Introduction} to {Privacy} {Engineering} and {Risk} {Management} in {Federal} {Systems}.
\newblock Technical Report NIST Internal or Interagency Report (NISTIR) 8062, National Institute of Standards and Technology, January 2017.

\bibitem{leventhal1980}
Gerald~S Leventhal.
\newblock What should be done with equity theory? {New} approaches to the study of fairness in social relationships.
\newblock In {\em Social exchange: {Advances} in theory and research}, pages 27--55. Springer, 1980.

\bibitem{gilliland_perceived_1993}
Stephen~W. Gilliland.
\newblock The {Perceived} {Fairness} of {Selection} {Systems}: {An} {Organizational} {Justice} {Perspective}.
\newblock {\em The Academy of Management Review}, 18(4):694--734, 1993.
\newblock Publisher: Academy of Management.

\bibitem{morse_ends_2022}
Lily Morse, Mike Horia~M. Teodorescu, Yazeed Awwad, and Gerald~C. Kane.
\newblock Do the {Ends} {Justify} the {Means}? {Variation} in the {Distributive} and {Procedural} {Fairness} of {Machine} {Learning} {Algorithms}.
\newblock {\em Journal of Business Ethics}, 181(4):1083--1095, December 2022.

\bibitem{grgic-hlaca_human_2018}
Nina Grgic-Hlaca, Elissa~M. Redmiles, Krishna~P. Gummadi, and Adrian Weller.
\newblock Human {Perceptions} of {Fairness} in {Algorithmic} {Decision} {Making}: {A} {Case} {Study} of {Criminal} {Risk} {Prediction}.
\newblock In {\em Proceedings of the 2018 {World} {Wide} {Web} {Conference} on {World} {Wide} {Web} - {WWW} '18}, pages 903--912, Lyon, France, 2018. ACM Press.

\bibitem{crawford2014bigDataDueProcess}
Kate Crawford and Jason Schultz.
\newblock Big data and due process: {Toward} a framework to redress predictive privacy harms.
\newblock {\em BCL Rev.}, 55:93, 2014.
\newblock Publisher: HeinOnline.

\bibitem{predicting_performance_waheed_2020}
Hajra Waheed, Saeed-Ul Hassan, Naif~Radi Aljohani, Julie Hardman, Salem Alelyani, and Raheel Nawaz.
\newblock Predicting academic performance of students from {VLE} big data using deep learning models.
\newblock {\em Computers in Human Behavior}, 104:106189, 2020.

\bibitem{kaurSystematicReviewPrediction-2020}
Puninder Kaur, Amandeep Kaur, and Rajwinder Kaur.
\newblock A {Systematic} {Review} {About} {Prediction} of {Academic} {Behavior} {Through} {Data} {Mining} {Techniques}.
\newblock {\em Journal of Computational and Theoretical Nanoscience}, 17(11):5162--5166, November 2020.

\bibitem{tiongEcheatingPreventionMeasures-2021}
Leslie Ching~Ow Tiong and HeeJeong~Jasmine Lee.
\newblock E-cheating {Prevention} {Measures}: {Detection} of {Cheating} at {Online} {Examinations} {Using} {Deep} {Learning} {Approach} -- {A} {Case} {Study}, January 2021.
\newblock arXiv:2101.09841 [cs].

\bibitem{gaba_predictive_2024}
Neha Gaba.
\newblock Predictive {Analytics} in {Hiring}: {The} {Key} to {Forecasting} {Workforce} {Needs} - {Blog}, April 2024.

\bibitem{punnoose_prediction_2016}
Rohit Punnoose and Pankaj Ajit.
\newblock Prediction of {Employee} {Turnover} in {Organizations} using {Machine} {Learning} {Algorithms}.
\newblock In {\em International {Journal} of {Advanced} {Research} in {Artificial} {Intelligence}}, volume~5, 2016.
\newblock ISSN: 21654069, 21654050 Issue: 9 Journal Abbreviation: ijarai.

\bibitem{lacroux_should_2022}
Alain Lacroux and Christelle Martin-Lacroux.
\newblock Should {I} {Trust} the {Artificial} {Intelligence} to {Recruit}? {Recruiters}’ {Perceptions} and {Behavior} {When} {Faced} {With} {Algorithm}-{Based} {Recommendation} {Systems} {During} {Resume} {Screening}.
\newblock {\em Frontiers in Psychology}, 13, July 2022.
\newblock Publisher: Frontiers.

\bibitem{cassidy2024}
Nazanin Cassidy, Kat.
\newblock U.s. job-seekers’ organizational justice perceptions of emotion ai-enabled interviews.
\newblock {\em Proceedings of the ACM on Human-Computer Interaction}, 2024.

\bibitem{Alene2022}
Kelsey Alene, Hilke Lauren, Paul Mona, and Julia.
\newblock Resume format, linkedin urls and other unexpected influences on ai personality prediction in hiring: Results of an audit.
\newblock {\em Proceedings of the 2022 AAAI/ACM Conference on AI, Ethics, and Society}, 2022.

\bibitem{edtech-pets22}
Rakibul Hasan and Mario Fritz.
\newblock Understanding {Utility} and {Privacy} of {Demographic} {Data} in {Education} {Technology} by {Causal} {Analysis} and {Adversarial}-{Censoring}.
\newblock {\em Proceedings on Privacy Enhancing Technologies}, 2022(2):245--262, April 2022.

\bibitem{overlearning}
Congzheng Song and Vitaly Shmatikov.
\newblock Overlearning reveals sensitive attributes.
\newblock {\em arXiv preprint arXiv:1905.11742}, 2019.

\bibitem{hendersonSelfDestructingModelsIncreasing-2023}
Peter Henderson, Eric Mitchell, Christopher Manning, Dan Jurafsky, and Chelsea Finn.
\newblock Self-{Destructing} {Models}: {Increasing} the {Costs} of {Harmful} {Dual} {Uses} of {Foundation} {Models}.
\newblock In {\em Proceedings of the 2023 {AAAI}/{ACM} {Conference} on {AI}, {Ethics}, and {Society}}, {AIES} '23, pages 287--296, New York, NY, USA, August 2023. Association for Computing Machinery.

\bibitem{wilsonGenderRaceIntersectional-2024}
Kyra Wilson and Aylin Caliskan.
\newblock Gender, {Race}, and {Intersectional} {Bias} in {Resume} {Screening} via {Language} {Model} {Retrieval}.
\newblock {\em Proceedings of the AAAI/ACM Conference on AI, Ethics, and Society}, 7(1):1578--1590, October 2024.
\newblock Number: 1.

\bibitem{obermeyerDissectingRacialBiasHealth2019}
Ziad Obermeyer, Brian Powers, Christine Vogeli, and Sendhil Mullainathan.
\newblock Dissecting racial bias in an algorithm used to manage the health of populations.
\newblock {\em Science}, 366(6464):447--453, October 2019.

\bibitem{magassaInclusiveJusticeApplying-2024}
Lassana Magassa and Batya Friedman.
\newblock Toward inclusive justice: {Applying} the {Diverse} {Voices} design method to improve the {Washington} {State} {Access} to {Justice} {Technology} {Principles}.
\newblock {\em ACM J. Responsib. Comput.}, 1(3):18:1--18:30, July 2024.

\bibitem{edm_and_privacy}
Mark Klose, Vasvi Desai, Yang Song, and Edward Gehringer.
\newblock {EDM} and privacy: {Ethics} and legalities of data collection, usage, and storage.
\newblock {\em International Educational Data Mining Society}, 2020.
\newblock Publisher: ERIC.

\bibitem{using_demographics_in_edm_2020}
Luc Paquette, Jaclyn Ocumpaugh, Ziyue Li, Alexandra Andres, and Ryan Baker.
\newblock Who's learning? {Using} demographics in {EDM} research.
\newblock {\em Journal of Educational Data Mining}, 12(3):1--30, 2020.
\newblock Publisher: ERIC.

\bibitem{big-five-original}
Lewis~R Goldberg.
\newblock The development of markers for the {Big}-{Five} factor structure.
\newblock {\em Psychological assessment}, 4(1):26, 1992.
\newblock Publisher: American Psychological Association.

\bibitem{hakimiRelationshipsPersonalityTraits-2011}
Soraya Hakimi, Elaheh Hejazi, and Masoud~Gholamali Lavasani.
\newblock The {Relationships} {Between} {Personality} {Traits} and {Students}’ {Academic} {Achievement}.
\newblock {\em Procedia - Social and Behavioral Sciences}, 29:836--845, January 2011.

\bibitem{timmons_pre-employment_2020}
Kelly~Cahill Timmons.
\newblock Pre-{Employment} {Personality} {Tests}, {Algorithmic} {Bias}, and the {Americans} with {Disabilities} {Act}.
\newblock {\em Penn State Law Review}, 125:389, 2020.

\bibitem{kamble_innovative_2022}
Priyanka Kamble and Umesh Kulkarni.
\newblock An {Innovative} {Approach} of {Personality} {Recognition} for {E}-{Recruitment}.
\newblock In {\em 2022 5th {International} {Conference} on {Advances} in {Science} and {Technology} ({ICAST})}, pages 324--328, December 2022.

\bibitem{van_mil_promises_2021}
Frenk~C.J. Van~Mil, Ayushi Rastogi, and Andy Zaidman.
\newblock Promises and {Perils} of {Inferring} {Personality} on {GitHub}.
\newblock In {\em Proceedings of the 15th {ACM} / {IEEE} {International} {Symposium} on {Empirical} {Software} {Engineering} and {Measurement} ({ESEM})}, pages 1--11, Bari Italy, October 2021. ACM.

\bibitem{giritlioglu_multimodal_2021}
Dersu Giritlio\u{g}lu, Burak Mandira, Selim~Firat Yilmaz, Can~Ufuk Ertenli, Berhan~Faruk Akg\"ur, Merve K{\i}n{\i}kl{\i}o\u{g}lu, Asl{\i}~G\"ul Kurt, Emre Mutlu, \c{S}eref~Can G\"urel, and Hamdi Dibeklio\u{g}lu.
\newblock Multimodal analysis of personality traits on videos of self-presentation and induced behavior.
\newblock {\em Journal on Multimodal User Interfaces}, 15(4):337--358, December 2021.

\bibitem{roemmich_emotion_2023}
Kat Roemmich, Florian Schaub, and Nazanin Andalibi.
\newblock Emotion {AI} at {Work}: {Implications} for {Workplace} {Surveillance}, {Emotional} {Labor}, and {Emotional} {Privacy}.
\newblock In {\em Proceedings of the 2023 {CHI} {Conference} on {Human} {Factors} in {Computing} {Systems}}, pages 1--20, Hamburg Germany, April 2023. ACM.

\bibitem{roemmich_values_2023}
Kat Roemmich, Tillie Rosenberg, Serena Fan, and Nazanin Andalibi.
\newblock Values in {Emotion} {Artificial} {Intelligence} {Hiring} {Services}: {Technosolutions} to {Organizational} {Problems}.
\newblock {\em Proceedings of the ACM on Human-Computer Interaction}, 7(CSCW1):1--28, April 2023.

\bibitem{ligeiro2024recruitment}
Nuno Ligeiro, Ivo Dias, and Ana Moreira.
\newblock Recruitment and selection process using artificial intelligence: How do candidates react?
\newblock {\em Administrative Sciences}, 14(7):155, 2024.
\newblock Accessed: 2025-04-10.

\bibitem{marrone2022creativity}
Rebecca Marrone, Victoria Taddeo, and Gillian Hill.
\newblock Creativity and artificial intelligence—a student perspective.
\newblock {\em Journal of Intelligence}, 10(3):65, 2022.
\newblock Accessed: 2025-04-10.

\bibitem{nugent2020recruitment}
Selin~E. Nugent, Paul Jackson, Susan Scott-Parker, James Partridge, Rebecca Raper, Alex Shepherd, Chara Bakalis, Arijit Mitra, Jintao Long, Kevin Maynard, and Nigel Crook.
\newblock Recruitment ai has a disability problem: Questions employers should be asking to ensure fairness in recruitment.
\newblock Technical report, Institute for Ethical Artificial Intelligence, Oxford Brookes University, 2020.
\newblock Accessed: 2025-04-10.

\bibitem{buyl2022tackling}
Maarten Buyl, Christina Cociancig, Cristina Frattone, and Nele Roekens.
\newblock Tackling algorithmic disability discrimination in the hiring process: An ethical, legal and technical analysis.
\newblock In {\em Proceedings of the 2022 ACM Conference on Fairness, Accountability, and Transparency (FAccT)}, 2022.
\newblock Accessed: 2025-04-10.

\bibitem{caloBoundariesPrivacyHarm-2011}
Ryan Calo.
\newblock The {Boundaries} of {Privacy} {Harm}.
\newblock {\em Indiana Law Journal}, 86:1131, 2011.

\bibitem{abercrombie_collaborative_2024}
Gavin Abercrombie, Djalel Benbouzid, Paolo Giudici, Delaram Golpayegani, Julio Hernandez, Pierre Noro, Harshvardhan Pandit, Eva Paraschou, Charlie Pownall, Jyoti Prajapati, Mark~A. Sayre, Ushnish Sengupta, Arthit Suriyawongkul, Ruby Thelot, Sofia Vei, and Laura Waltersdorfer.
\newblock A {Collaborative}, {Human}-{Centred} {Taxonomy} of {AI}, {Algorithmic}, and {Automation} {Harms}, July 2024.
\newblock arXiv:2407.01294 [cs].

\bibitem{gdpr}
General data protection regulation.
\newblock \url{https://gdpr-info.eu/}, 2024.

\bibitem{barocas2016bigdata}
Solon Barocas and Andrew~D. Selbst.
\newblock Big data’s disparate impact.
\newblock {\em California Law Review}, 104(3):671--732, 2016.

\bibitem{wachter2019reasonable}
Sandra Wachter and Brent Mittelstadt.
\newblock A right to reasonable inferences: Re-thinking data protection law in the age of big data.
\newblock {\em Columbia Business Law Review}, 2019(2):494--620, 2019.

\bibitem{ahmad2023privacy}
T.~Ahmad, N.~Ismail, and N.~A. Aziz.
\newblock Implications of artificial intelligence on student privacy and autonomy.
\newblock {\em Education and Information Technologies}, 28:10241--10260, 2023.

\bibitem{susser2019manipulation}
Daniel Susser, Beate Roessler, and Helen Nissenbaum.
\newblock Technology, autonomy, and manipulation.
\newblock {\em Internet Policy Review}, 8(2), 2019.

\bibitem{caliskan2017semantics}
Aylin Caliskan, Joanna~J. Bryson, and Arvind Narayanan.
\newblock Semantics derived automatically from language corpora contain human-like biases.
\newblock {\em Science}, 356(6334):183--186, 2017.

\bibitem{akgun2021ethics}
S.~Akgun and C.~Greenhow.
\newblock Artificial intelligence in education: Addressing ethical challenges in k-12 settings.
\newblock {\em British Journal of Educational Technology}, 52(4):1541--1558, 2021.

\bibitem{martinPredatoryPredictionsEthics-2023}
Kirsten Martin.
\newblock Predatory predictions and the ethics of predictive analytics.
\newblock {\em Journal of the Association for Information Science and Technology}, 74(5):531--545, 2023.
\newblock \_eprint: https://onlinelibrary.wiley.com/doi/pdf/10.1002/asi.24743.

\bibitem{hanson2021ofqual}
Janet Hanson.
\newblock Learning from ofqual’s 2020 grading algorithm failure.
\newblock {\em British Educational Research Journal}, 47(3):729--737, 2021.

\bibitem{tenzer2023defamation}
Lisa Y.~Garfield Tenzer.
\newblock Defamation in the age of artificial intelligence.
\newblock {\em SSRN Electronic Journal}, 2023.
\newblock Available at SSRN: https://ssrn.com/abstract=4567891.

\bibitem{brundage2018malicious}
Miles Brundage, Shahar Avin, Jack Clark, Helen Toner, et~al.
\newblock The malicious use of artificial intelligence: Forecasting, prevention, and mitigation.
\newblock {\em arXiv preprint arXiv:1802.07228}, 2018.

\bibitem{apa2023stress}
{American Psychological Association}.
\newblock Worries about artificial intelligence, surveillance at work may be connected to poor mental health, 2023.
\newblock Press Release, Sep. 7, 2023.

\bibitem{prolific}
Prolific · {Quickly} find research participants you can trust.

\bibitem{tavakol_making_2011}
Mohsen Tavakol and Reg Dennick.
\newblock Making sense of {Cronbach}'s alpha.
\newblock {\em International Journal of Medical Education}, 2:53--55, June 2011.

\bibitem{field2012discoveringStatR}
Andy Field, Zoe Field, and Jeremy Miles.
\newblock Discovering statistics using {R}.
\newblock 2012.
\newblock Publisher: sage.

\bibitem{prince2023privacy}
Jeffrey Prince and Scott Wallsten.
\newblock Privacy preferences differ by gender and age, but not by income.
\newblock {\em The Technology Policy Institute}, 2023.
\newblock Accessed: 2025-04-09.

\bibitem{pierson2017gender}
Emma Pierson.
\newblock Gender differences in beliefs about algorithmic fairness.
\newblock {\em arXiv preprint arXiv:1712.09124}, 2017.
\newblock Accessed: 2025-04-10.

\bibitem{make6010017}
Ekaterina Novozhilova, Kate Mays, Sejin Paik, and James~E. Katz.
\newblock More capable, less benevolent: Trust perceptions of ai systems across societal contexts.
\newblock {\em Machine Learning and Knowledge Extraction}, 6(1):342--366, 2024.

\bibitem{diangelo2022whitefragility}
Robin DiAngelo.
\newblock {\em White fragility: {Why} {Understanding} racism can be so hard for white people (adapted for young adults)}.
\newblock Beacon press, 2022.

\bibitem{guPrivacyAccuracyModel-2022}
Xiuting Gu, Zhu Tianqing, Jie Li, Tao Zhang, Wei Ren, and Kim-Kwang~Raymond Choo.
\newblock Privacy, accuracy, and model fairness trade-offs in federated learning.
\newblock {\em Computers \& Security}, 122:102907, November 2022.

\bibitem{changPrivacyRisksAlgorithmic-2021}
Hongyan Chang and Reza Shokri.
\newblock On the {Privacy} {Risks} of {Algorithmic} {Fairness}.
\newblock In {\em 2021 {IEEE} {European} {Symposium} on {Security} and {Privacy} ({EuroS}\&{P})}, pages 292--303, September 2021.

\bibitem{massey_requirements-based_2008}
Aaron~K. Massey and Annie~I. Antón.
\newblock A {Requirements}-based {Comparison} of {Privacy} {Taxonomies}.
\newblock In {\em 2008 {Requirements} {Engineering} and {Law}}, pages 1--5, September 2008.

\bibitem{landuytPrivacyImpactTree-2025}
Dimitri~Van Landuyt.
\newblock Privacy {Impact} {Tree} {Analysis} ({PITA}): {A} {Tree}-based {Privacy} {Threat} {Modeling} {Approach}.
\newblock {\em IEEE Transactions on Software Engineering}, pages 1--23, 2025.

\bibitem{kelso_trust_2024}
Easton Kelso, Ananta Soneji, Sazzadur Rahaman, Yan Shoshitaishvili, and Rakibul Hasan.
\newblock Trust, {Because} {You} {Can}'t {Verify}: {Privacy} and {Security} {Hurdles} in {Education} {Technology} {Acquisition} {Practices}.
\newblock In {\em Proceedings of the 2024 on {ACM} {SIGSAC} {Conference} on {Computer} and {Communications} {Security}}, {CCS} '24, pages 1656--1670, New York, NY, USA, December 2024. Association for Computing Machinery.

\bibitem{colnago-concern-preference}
Jessica Colnago, Lorrie~Faith Cranor, Alessandro Acquisti, and Kate~Hazel Jain.
\newblock Is it a {Concern} or a {Preference}? {An} {Investigation} into the {Ability} of {Privacy} {Scales} to {Capture} and {Distinguish} {Granular} {Privacy} {Constructs}.
\newblock In {\em Eighteenth symposium on usable privacy and security ({SOUPS} 2022)}, pages 331--346, Boston, MA, August 2022. USENIX Association.

\end{thebibliography}
